\documentclass[usenatbib]{mnras}
\usepackage{graphicx}
\usepackage{float}
\usepackage{amsmath}
\usepackage{amssymb}
\usepackage{textcomp}
\usepackage{multirow}  
\usepackage[caption=false]{subfig}
\usepackage{color}
\usepackage{gensymb}   
\usepackage{hyperref} 
\usepackage{xspace} 
\usepackage{afterpage} 

\newcommand{\comment}[1]{\ignorespaces}

\makeatletter
\newcommand\footnoteref[1]{\protected@xdef\@thefnmark{\ref{#1}}\@footnotemark}
\makeatother

\title[Satellites of the Nearest Hosts in LBT-SONG]{A Search for Satellite Galaxies of Nearby Star-Forming Galaxies with Resolved Stars in LBT-SONG}

\author[C. T. Garling et al.]{Christopher T. Garling,$^{1}$\thanks{E-mail: garling.14@osu.edu}
Annika H. G. Peter,$^2$
Christopher S. Kochanek,$^1$ \newauthor
David J. Sand,$^3$
Denija Crnojevi\'c$^4$
\\
$^1$CCAPP and Department of Astronomy, The Ohio State University, 140 W. 18th Ave., Columbus, OH 43210, USA \\
$^2$CCAPP, Department of Physics, and Department of Astronomy, The Ohio State University, 191 W. Woodruff Ave., Columbus, OH 43210, USA \\
$^3$Department of Astronomy and Steward Observatory, University of Arizona, 933 N. Cherry Avenue, Tucson, AZ 85719, USA \\
$^4$University of Tampa, 401 West Kennedy Boulevard, Tampa, FL 33606, USA \\
}

\date{Accepted XXX. Received YYY; in original form ZZZ}

\pubyear{2019}

\begin{document}
\label{firstpage}
\pagerange{\pageref{firstpage}--\pageref{lastpage}}
\maketitle

\begin{abstract}
  We present results from a resolved stellar population search for dwarf satellite galaxies of six nearby (D $<5$ Mpc), sub-Milky-Way mass hosts using deep ($m\sim27$ mag) optical imaging from the Large Binocular Telescope. We perform image simulations to quantify our detection efficiency for dwarfs over a large range in luminosity and size, and develop a fast catalog-based emulator that includes a treatment of unresolved photometric blending. We discover no new dwarf satellites, but we recover two previously known dwarfs (DDO 113 and LV J1228+4358) with $M_{\text{V}}<-12$ that lie in our survey volume. We preview a new theoretical framework to predict satellite luminosity functions using analytic probability distribution functions and apply it to our sample, finding that we predict one fewer classical dwarf and one more faint dwarf ($M_{\text{V}}\sim-7.5$) than we find in our observational sample (i.e., the observational sample is slightly top-heavy). However, the overall number of dwarfs in the observational sample (2) is in good agreement with the theoretical expectations. Interestingly, DDO 113 shows signs of environmental quenching and LV J1228+4358 is tidally disrupting, suggesting that low-mass hosts may affect their satellites more severely than previously believed.
\end{abstract}

\begin{keywords} 
galaxies: dwarf -- galaxies: evolution
\end{keywords}

\section{Introduction}
Dwarf satellite galaxies are powerful cosmological probes on non-linear scales where there are few competing constraints, as these dwarf galaxies trace the underlying dark matter subhalo population \citep[e.g.,][]{Zheng2005,Zheng2007,Navarro2018}. In concert with galaxy evolution models for how galaxies inhabit halos, observations of satellite luminosity functions (SLFs) can be compared to predictions from various cosmologies to place constraints on alternate dark matter models or look for deviations from the prevailing $\Lambda$CDM cosmology \citep[e.g.,][]{Nierenberg2012,Kennedy2014,Nierenberg2016,Chau2017,Dooley2017a,Lovell2017,Jethwa2018,Kim2018a,Gilman2020,Lovell2020,Nadler2020,Newton2020,Safarzadeh2020,Nadler2021}. \par

However, large uncertainties in the galaxy evolution models can make these comparisons difficult, especially as the cosmological signal (principally, the subhalo mass function, or SMF) is degenerate with the stellar-mass-halo-mass (SMHM) relation at fixed host mass; both can fundamentally alter the shape and normalization of the SLF. This degeneracy can be broken by sampling hosts of different masses. The SMF at infall (i.e., the unevolved subhalo mass function) in $\Lambda$CDM scales universally with host mass, which is often described as the self-similarity of substructure in $\Lambda$CDM \citep[e.g.,][]{Gao2004,Gao2012,Jiang2014,Jiang2016}. This is not true of the evolved ($z=0$) SMF, which is affected by the tidal stripping of subhalos and is not strictly self-similar. However, simulations show the present-day stellar masses of dwarf satellites are closely correlated to their peak historical halo masses, which are typically their halo masses at first infall to their hosts \citep[e.g.,][]{Reddick2013,RodriguezPuebla2017,Buck2019c,Campbell2018,Behroozi2019,Moster2020,Wang2021}. Thus, the self-similarity of substructure in $\Lambda$CDM is imprinted on the SLF. The degeneracy between the SMHM function (galaxy evolution) and the SMF (cosmology) can be broken by obtaining SLFs of hosts with a range of halo masses, as the SMHM relation will be universal across all hosts, as will the shape of the SMF, while the normalization of the SMF is dependent on the host mass \citep[see, e.g., the discussion in \S8.1 of][]{Roberts2021}. Therefore, a sample of SLFs from hosts of varied masses can be modelled simultaneously to constrain both the SMHM relation and the SMF. \par

Thus far, SLFs have mostly been measured for galaxies around the mass of the Milky Way (MW), as they can be easily compared to the well-studied MW satellite population, and there is a low likelihood of such galaxies hosting no detectable dwarf galaxies \citep[e.g.,][]{Sand2014,Danieli2017,Geha2017,Mueller2017,Smercina2018,Crnojevic2019,Mueller2019,Bennet2020,Mao2020,Carlsten2021}. Studies at roughly constant host mass are good for understanding halo-to-halo scatter and the environmental dependence of SLFs but they are not optimal for understanding the SMF or the SMHM relation. Additionally, most surveys aim simply to discover new dwarfs without modelling the completeness of the search; these completeness corrections are essential to making comparisons with theoretical expectations. \par

An attractive approach for extending the range of host masses with measured SLFs is to go to lower host masses, as there are many nearby low-mass galaxies that can be surveyed. The dwarf galaxies associated with the Large Magellanic Cloud (LMC) found by the Dark Energy Survey \citep{TheDESCollaboration2015,TheDESCollaboration2015a,Nadler2020} are a first example that low-mass galaxies host significant satellite populations, as predicted by the self-similarity of substructure in $\Lambda$CDM. Success has been found in this regime by the MADCASH project, which recently discovered new dwarf satellites of the low-mass hosts NGC 2403 \citep{Carlin2016} and NGC 4214 \citep{Carlin2021}. The first results from the LBT-SONG survey identified two dwarf satellite candidates of NGC 628 \citep{Davis2021}. Additionally, \cite{Mueller2018a} and \cite{Mueller2020} assembled dwarf candidate catalogs for the diffuse Leo-I and Sculptor groups, which, if confirmed, will shed light on the SLF of dwarfs in group environments, and comparison to isolated systems may reveal environmental effects on the SLF. However, larger host samples are needed to constrain the SMF and SMHM relation.\par

Towards this end, we present a search for dwarf satellite galaxies of six nearby (3--4.5 Mpc), sub-MW-mass hosts using imaging data from the Large Binocular Telescope (LBT; \citealt{Hill2010a}) using the Large Binocular Camera (LBC; \citealt{Ragazzoni2006,Speziali2008}). These data are deep, achieving 50\% point-source completeness at $\sim27$ magnitude in the \emph{B}, \emph{V}, and \emph{R} bands, enabling us to detect dwarf satellites as faint as $M_{\text{V}} \, \sim -7$ (or $\text{M}_*\sim5\times10^4$ M$_{\odot}$) at roughly 50\% completeness by searching for dwarfs with resolved stellar populations. This sensitivity, which is better than most previous extragalactic surveys, indicates that we should be able to detect dwarfs at the transition between the classical and ultra-faint regimes. We determine our dwarf completeness function using both traditional image simulations and a novel catalog modelling approach that explicitly includes the effects of photometric blending.\par

A limitation of these data relative to other SLF-focused surveys is our restricted field of view. With only a single LBC pointing per host, we generally cover only the central 400--600 kpc$^2$ for our hosts, while some other surveys span the entire virial volume (typically $\geq$ 1 Mpc$^2$ in projection). Due to our limited volume, interpretation of our results depends heavily on the assumed radial distribution of satellites. However, theoretical predictions for the radial distributions of satellites are uncertain due to tidal disruption. Resolution studies of tidal stripping have shown that satellites are artificially disrupted at resolutions typical of cosmological simulations, while analogous satellites survive much longer at higher resolution \citep{Penarrubia2008,Penarrubia2010,Bosch2018,Bosch2018a}; these objects are called orphaned subhalos, and properly accounting for them is critical for many applications \citep[e.g., constructing semi-analytic models of galaxy evolution and fitting the SMHM relation and the MW satellite population;][]{Simha2017,Drakos2020,Jiang2021,Delfino2021,Moster2013,Behroozi2019,Nadler2019a}. The radial distribution is particularly susceptible to the effects of artificial disruption, as the tidal mass loss rate is a strong function of radius. When including orphaned subhalos, MW-scale simulations suggest that the subhalo radial distribution is even more centrally-concentrated than a cuspy dark matter host halo due to dynamical friction on the infalling satellites \citep[e.g.,][]{Han2016}. Since the strength of dynamical friction scales with the satellite halo mass \citep[e.g.,][]{BoylanKolchin2008}, the radial distribution for high-mass subhalos is more centrally-concentrated than for low-mass subhalos \citep[e.g., right panel of Figure 3 in][]{Han2016}. Thus, the central region should have the highest surface density of satellites, and should also host many of the most massive, and easily detectable, satellites.\par

Most environmental quenching mechanisms are also more effective at small host-centric radii. However, the environmental processes invoked to explain the quiescent (i.e., quenched) satellite galaxies of the Milky Way, such as accretion-shocked hot gas halos \citep[e.g.,][]{Birnboim2003,Keres2009} and ram pressure stripping \citep[e.g.,][]{Gunn1972,Mayer2006,Fillingham2016,Tonnesen2019,Stern2020}, have not been explored in detail for the lower host masses we examine here. Looking for dwarf satellite galaxies at small projected radii may then shed light on the quenching mechanisms active in sub-MW mass hosts. Using these LBT data, \cite{Garling2020} showed that the Fornax-analog DDO 113, which ceased forming stars less than 1 Gyr ago \citep{Weisz2011}, is undisrupted and was likely quenched due to the cessation of cold gas inflows upon entering the halo of its host, the LMC-analog NGC 4214. Similarly, \cite{Carlin2019} showed that DDO 44, a dwarf satellite of similar luminosity that also quenched about a Gyr ago \citep{Karachentsev2007,Weisz2011}, was also tidally disrupted by its $\sim2\times$LMC-mass host, NGC 2403. These results are somewhat surprising when compared to the SAGA survey, which focuses on MW-mass hosts and has found predominantly star-forming dwarf satellite galaxies \citep{Geha2017,Mao2020}. While the SAGA survey is quite different in methodology and host selection, the differences in their quenched fractions make it clear that environmental quenching mechanisms are not yet well-understood.\par

\begin{table*}
  \caption{Summary of observations and assumed host properties}

  \begin{tabular}{c c c c c c c c c}
    \hline
    \hline
    Host & R.A. & Decl. & Distance &  M$_*$\textsuperscript{a} & Exposure\textsuperscript{b} & 50\% complete\textsuperscript{b} & Subtended area & E(B$-$V)\textsuperscript{c} \\
    & [hms] & [dms] & [Mpc] & [log$_{10}$ (M$_{\odot}$)] & [s] & [mag] & [kpc$^2$] & [mag] \\
    \hline
    NGC 4214 & 12\textsuperscript{h}15\textsuperscript{m}39\textsuperscript{s} & $+36\degree19^{\prime}37^{\prime\prime}$ & 3.04\textsuperscript{d} & $8.55\pm 0.10$ & 1800, 1800, 5400 & 26.8, 26.6, 26.9 & 400 & 0.0188\\
    IC 2574 & 10\textsuperscript{h}28\textsuperscript{m}23\textsuperscript{s} & $+68\degree24^{\prime}44^{\prime\prime}$ & 3.81\textsuperscript{d} & $8.72\pm0.14$ & 2040, 1920, 6000 & 27.0, 26.6, 26.9 & 500 & 0.0311 \\
    NGC 3077 & 10\textsuperscript{h}03\textsuperscript{m}19\textsuperscript{s} & $+68\degree44^{\prime}02^{\prime\prime}$ & 3.83\textsuperscript{d} & $9.17\pm 0.10$ & 2160, 2160, 5760 & 27.0, 26.7, 26.9 & 520 & 0.0576 \\
    NGC 4449 & 12\textsuperscript{h}28\textsuperscript{m}11\textsuperscript{s} & $+44\degree05^{\prime}37^{\prime\prime}$ & 4.02\textsuperscript{e} & $9.03\pm0.10$ & 3000, 3000, 8160 & 27.1, 26.7, 27.0 & 520 & 0.0166 \\
    NGC 4826 & 12\textsuperscript{h}56\textsuperscript{m}44\textsuperscript{s} & $+21\degree40^{\prime}59^{\prime\prime}$ & 4.40\textsuperscript{f} & $10.20\pm0.10$ & 6480, 6480, 18720 & 27.2, 26.9, 27.2 & 580 & 0.0356 \\
    NGC 4236 & 12\textsuperscript{h}16\textsuperscript{m}42\textsuperscript{s} & $+69\degree27^{\prime}45^{\prime\prime}$ & 4.45\textsuperscript{g} & $9.19\pm0.11$ & 4080, 4320, 10080 & 27.2, 26.8, 27.1 & 600 & 0.0125 \\
    \hline
  \end{tabular} \\
  \textsuperscript{a} \cite{Leroy2019}.
  \textsuperscript{b} The three values in these columns are for \emph{B}, \emph{V}, and \emph{R}, respectively.
  \textsuperscript{c} \cite{Schlegel1998, Schlafly2011}. \textsuperscript{d} \cite{Dalcanton2009}. \textsuperscript{e} \cite{Sabbi2018}; see also \cite{McQuinn2010}. \textsuperscript{f} \cite{Jacobs2009}. \textsuperscript{g} \cite{Karachentsev2002}.
  \label{Table:observations}
\end{table*}

In this work, we search for dwarf satellites of six sub-MW mass hosts using LBT/LBC imaging, carefully derive our detection efficiency, and compare the results to theoretical expectations. We outline the acquisition and processing of our imaging data in \S \ref{section:data}. We present our method to search the images for dwarf satellites using resolved stars in \S \ref{section:dwarfsearch}. In \S \ref{Section: selection} we compare two methods for deriving our observational selection function: a standard image simulation technique, and a novel catalog modelling approach that explicitly models the effects of photometric blending. In \S \ref{section:results} we present the results from our search and compare them to predictions from a new theoretical framework for modelling SLFs. 

\section{Observations and Data Reduction} \label{section:data}

Our observations were originally obtained in the search for failed supernovae described in \cite{Kochanek2008}, with results presented in \cite{Adams2017a}, \cite{Basinger2020}, and \cite{Neustadt2021}. This survey monitors 27 nearby, star-forming galaxies with the blue and red channels of the LBC \citep{Ragazzoni2006,Speziali2008} on the LBT \citep{Hill2010a}. The LBT has two 8.4m primary mirrors and the LBC consists of two prime focus cameras, with one optimized for blue light and one for red. Each camera uses four 2048 $\times$ 4608 pixel CCDs with a scale of $0\farcs225$ per pixel and a field of view of $23^{\prime} \times 23^{\prime}$. The earliest data we use here were obtained in 2009 and the latest were taken in 2017. Typically two epochs of data were obtained each year. Each epoch consisted of three \emph{R} band exposures on LBC red paired with single \emph{U}, \emph{B}, and \emph{V} band exposures on LBC blue. Individual exposure times were scaled to reach a fixed point-source luminosity for each galaxy \citep[see][]{Gerke2015} and are typically 2--4 minutes for the hosts studied here. Seeing varied from $0\farcs8$ to $1\farcs5$, but we only used images with seeing better than $1\farcs1$. We additionally exclude images with high sky brightness, which comprised roughly 5\% of the exposures. A summary of our observations is given in Table \ref{Table:observations}.\par

We reprocess the raw images for the seven closest galaxies and apply resolved star methods to search for dwarf satellite galaxies; we refer to this sample as the LBT-SONG near-sample. More distant galaxies in the sample of \cite{Kochanek2008} will be searched with integrated light methods by Davis et al. (2021, in preparation), which we will refer to as the LBT-SONG far-sample; the first results from this sample were presented in \cite{Davis2021} and revealed two dwarf satellite candidates of NGC 628. Our methods for data reduction, photometry, calibration, and artificial star tests follow closely those of \cite{Garling2020} so we abbreviate our description here, focusing mainly on changes from that analysis. \par

\begin{figure}
	\centering
        \includegraphics[width=0.45\textwidth,page=1]{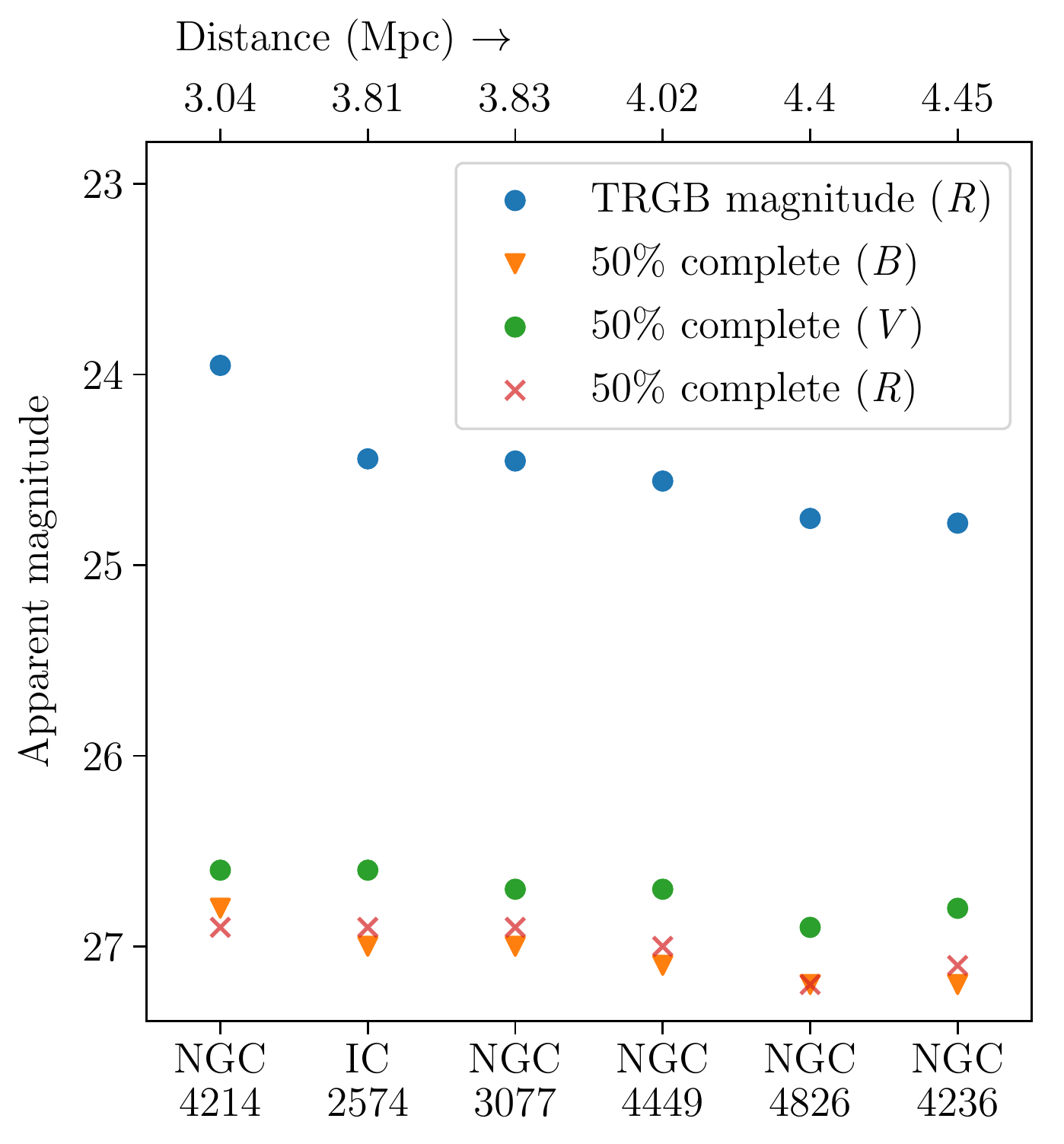} 
	\caption{50\% completeness values for \emph{B}, \emph{V}, and \emph{R} derived from artificial star tests compared to the tip of the red giant branch (TRGB; absolute magnitude in \emph{R} is $-3.462$) magnitude of the isochrone we use for our dwarf search. We are sensitive to $\sim$\,2  mags of the RGB even for the most distant hosts. Our 90\% completeness limits are about 1 mag brighter than our 50\% completeness limits.} 
	\label{figure:completeness_curve}
\end{figure}

\subsection{Data Reduction}\label{subsection:reduction}
The \textsc{iraf mscred} package was used to apply overscan correction, bias subtraction, and flat fielding as described in \cite{Gerke2015}.
An initial astrometric calibration was made using a local installation of the \texttt{astrometry.net} code base \citep{Lang2010} with custom astrometric reference files built from SDSS-DR13 \citep{Alam2015}. We then used \texttt{SCAMP} \citep{Bertin2006} to perform internal and external astrometric calibrations with focal plane distortion corrections, cross-matching the astrometry of our input images to standard deviations $\sim0\farcs1$ and calibrating to GAIA-DR2 \citep{GaiaCollaboration2018} sources with similar precision. \par

Following this astrometric calibration, we co-added our individual exposures using \texttt{SWarp} \citep{Bertin2002}. Individual images were resampled using a \textsc{lanczos3} interpolation function, and co-addition was performed with the clipped-mean algorithm from \cite{Gruen2014}. Most final co-adds have seeing between $1\farcs0 - 1\farcs2$ with a median of $1\farcs1$. \par

\subsection{Photometry and Calibration} \label{subsection:photometry}
We performed point spread function (PSF) fitting photometry using the \textsc{daophot}, \textsc{allstar}, and \textsc{allframe} packages \citep{Stetson1987,Stetson1994}. We ran \textsc{allstar} in two iterations, once on the original science image and again on the science image with the sources from the first pass subtracted to detect faint sources originally missed. We then performed forced photometry simultaneously in all bands for sources detected in at least two filters by \textsc{allstar} using \textsc{allframe} \citep{Stetson1994}. We remove poorly measured sources by eliminating stars with photometric errors greater than 0.4 mag or $\chi$, a measure of the goodness-of-fit of the model PSF, greater than 2 in the \textsc{allframe} catalogs. We use the \textsc{daophot} sharpness parameter as our statistic for separating stars and galaxies, which has been shown to work well \citep[e.g., Figure 4 in][]{Annunziatella2013}. We classify sources with absolute values of sharpness less than 2 as stars in our final catalog. \par

We calibrate our photometry by bootstrapping onto SDSS-DR13 \citep{Alam2015} and correcting for Galactic extinction. We used relations from \cite{Jordi2006} to convert SDSS magnitudes to \emph{U}, \emph{B}, \emph{V}, and \emph{R} with full error propagation. We match our photometric catalog to the SDSS-DR13 `stars' catalog and use the sources in common to fit zero points and color terms for all bands simultaneously, accounting for covariances between the zero points and color terms for our four bands. Mean calibration errors are $0.03-0.05$ mag, determined by 10,000 iterations of bootstrap resampling. For IC 2574 and NGC 4236, we calibrated instead to PanSTARRS-DR2 \citep{Chambers2016,Flewelling2016} due to the unavailability of SDSS-DR13 data in this regions. We use the \cite{Tonry2012} filter conversions for \emph{B}, \emph{V}, and \emph{R}, and the \cite{Karaali2005} conversion for \emph{U}, but the zeropoint and color term calibration methods remain the same. \par

Reported magnitudes are corrected for Galactic extinction with $E \, (B-V)$ values obtained by interpolating the \cite{Schlegel1998} dust maps with the updated scalings from \cite{Schlafly2011}. Hereafter, all quoted magnitudes have been corrected for Galactic extinction.

\subsection{Artificial Star Tests} \label{subsection:artificialstartests}
To measure our photometric errors and completeness as a function of magnitude and color, we perform artificial star tests on our final co-adds as in \cite{Garling2020}, excluding the central chips where the hosts take up a majority of the frame. \par 

We create a mock catalog of artificial stars on a spatial grid with a $\sim15^{\prime\prime}$ spacing to prevent artificial crowding. We use the \textsc{addstar} routine from \textsc{daophot} \citep{Stetson1987,Stetson1994} to insert the stars, which makes full use of the best-fit analytic form of the PSF, its empirical correction table, and the spatial variation of the PSF \citep[see][]{Stetson1992}. We create 10 artificial images per chip per band, each with 2,000 artificial stars for a total of 80,000 artificial stars per host. The artificial images are processed with the same photometric pipeline as our science images and we apply the same cuts in chi, sharpness, and magnitude error to the resulting photometric catalog. Our completeness curves for NGC 4214 are shown in Figure 1 of \cite{Garling2020}, and 50\% completeness values for \emph{B}, \emph{V}, and \emph{R} are given for each host in Table \ref{Table:observations} and shown in Figure \ref{figure:completeness_curve}. Our 90\% completeness values are generally 1 mag brighter than the 50\% completeness values. \emph{U} is excluded because we do not use it for candidate detection due to its $\sim$1 mag shallower depth and the expectation that most dwarfs we are sensitive to will not have young stars that are bright in \emph{U}. \par

\section{Dwarf Search Methods} \label{section:dwarfsearch}
To maximize the signal from potential dwarf galaxies and reduce contamination from foreground MW stars and background galaxies, we filter our point-source catalogs with a 10 Gyr, $\text{[Fe/H]}=-2$ dex isochrone from PARSEC \citep{Aringer2009,Bressan2012,Chen2014,Marigo2017}. We considered other isochrone choices, but found that using an old isochrone maximized the signal from potential candidates due to our ability to resolve the upper red giant branch (RGB). We use the filtering scheme from \cite{Garling2020}, where stars are Gaussian-weighted according to their proximity to the model isochrone and the photometric error from the artificial star tests. Weights range from $0 \leq w_i \leq 1$, indicating very poor and very good matches to the isochrone, respectively. We use both the $B-V$ and $V-R$ colors and we include only stars with weights greater than 0.2 to remove stars with low membership probabilities. \par

\begin{figure*}
  \centering
  \includegraphics[width=0.56\textwidth,page=1]{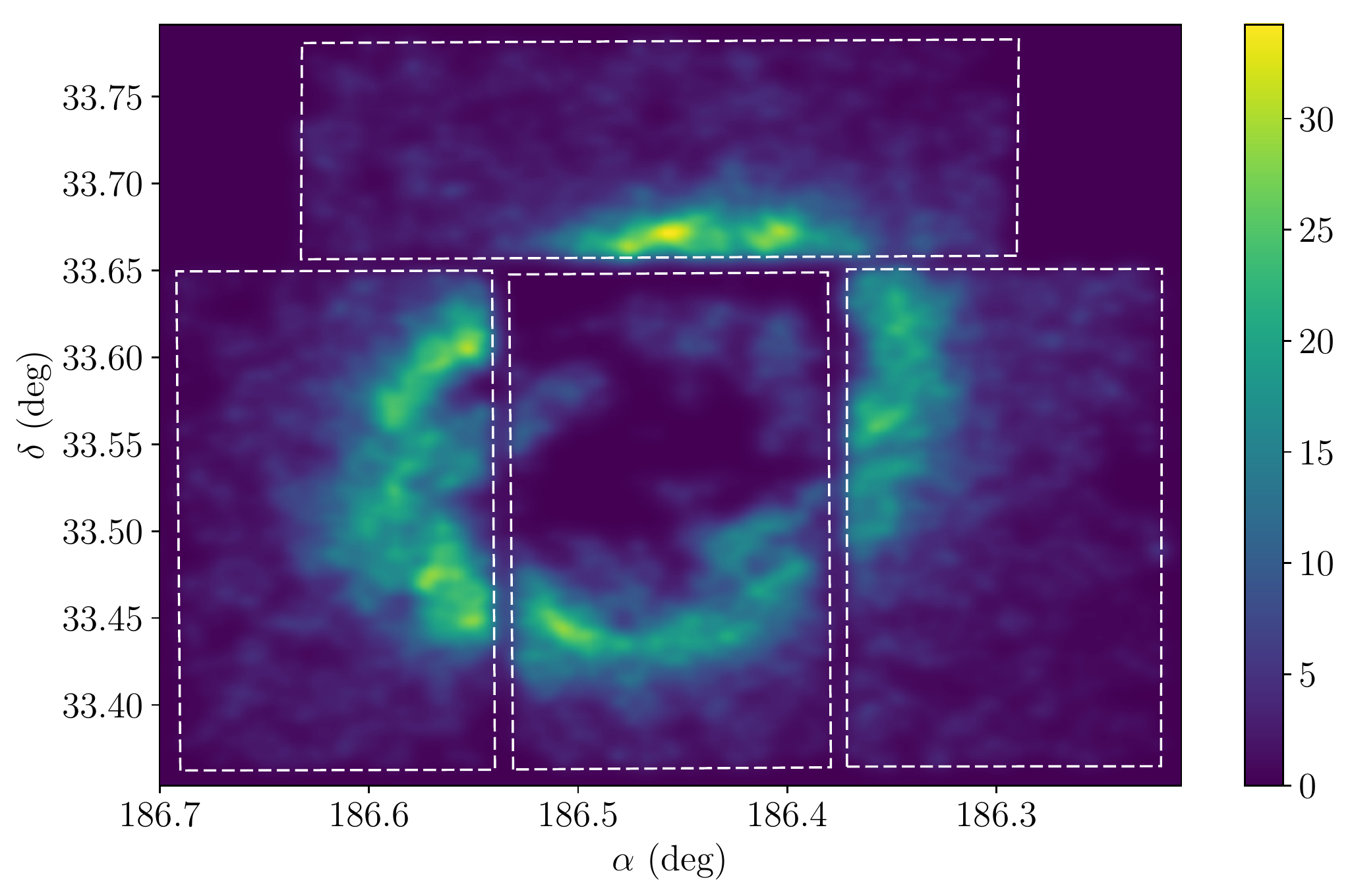} \includegraphics[width=0.43\textwidth,page=1]{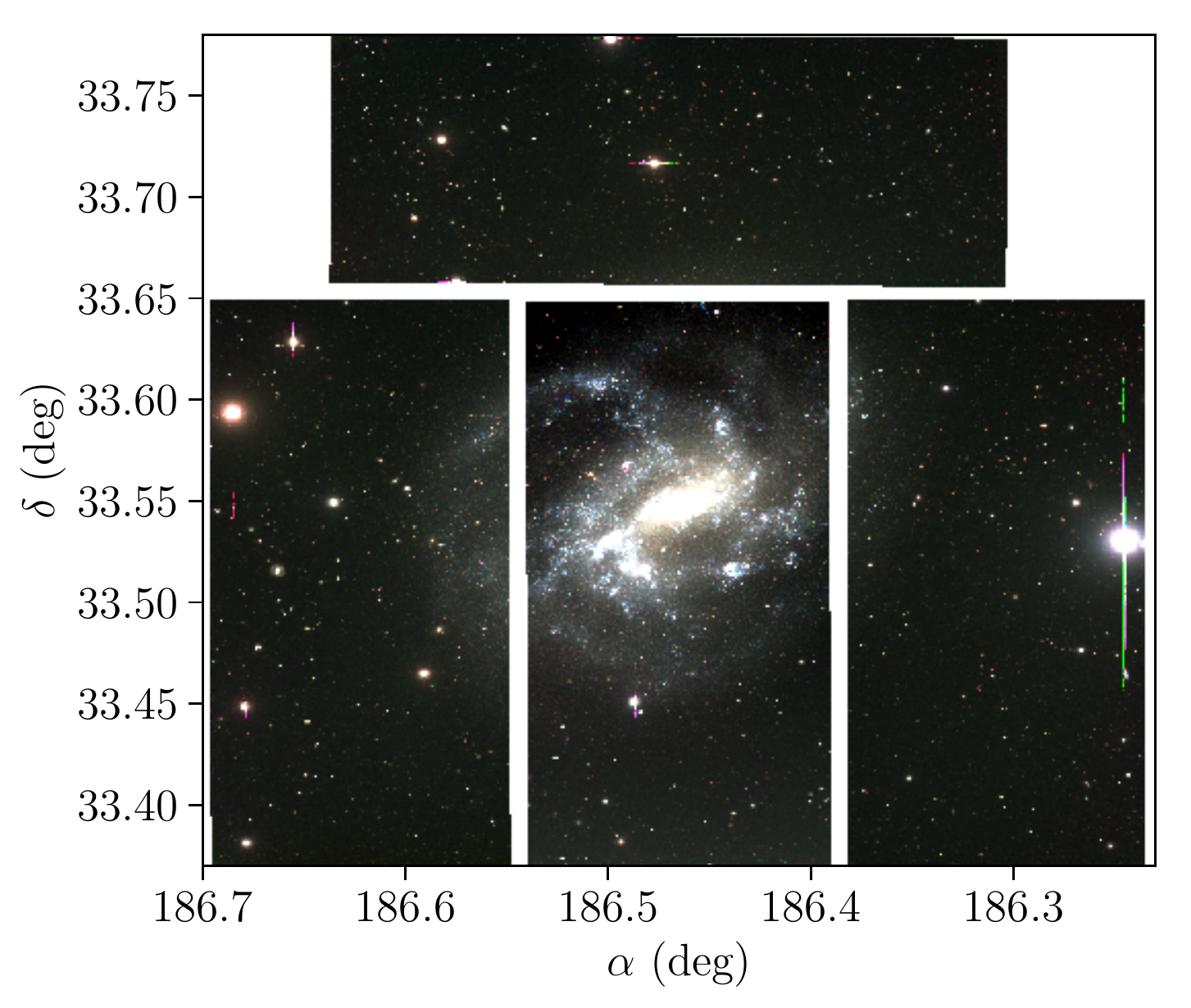}
  \caption{\emph{Left:} KDE map for NGC 4395 with a kernel size of 300 pc. The colorbar shows raw density estimates computed from Equation \ref{equation:kde}. The host lies in the central region, where our point-source photometry is poor. The bright, ring-like feature is due to stars from the host. The white, dashed boxes indicate the LBC CCD chips. \emph{Right:} Composite color (\emph{R, V, B} filters mapped to RGB) of our LBT/LBC images of NGC 4395.}
  \label{Figure:kdemap}
\end{figure*}

We then use these weights to construct a weighted kernel density estimation (KDE) using a Plummer \citep{Plummer1911} kernel because the Plummer model reproduces the surface brightness profiles of Milky Way satellites well \citep[e.g.,][]{Munoz2018}. The value of the KDE at position $\vec{x}$ is 

\begin{equation} \label{equation:kde}
  \hat{f}_h(\vec{x}) = \sum_{i=1}^{N} \ w_i \ \frac{h^2}{ \pi \left[h^2 + \left(\vec{x}-\vec{x}_i\right)^2 \right]^2}
\end{equation}

\noindent where the $\vec{x}_i$ are the positions of the stars, $N$ is the total number of stars, $h$ is the bandwidth of the kernel and equal to the Plummer scale radius, and the sum is over all stars $i=1...N$ within $3h$ of $\vec{x}$. 

We evaluate $\hat{f}_h(\vec{x})$ on an $800^2$ grid over the full field of view and use kernel bandwidths of 200, 300, 500, and 1000 pc at the assumed distances of our hosts to produce four maps that resemble the RGB maps common in other studies \citep[e.g.,][]{Carlin2019}. An example of a KDE map of $\hat{f}_h(\vec{x})$ with $h=300$ pc for NGC 4395 is shown in Figure \ref{Figure:kdemap}. The greatest difficulty we have to contend with is the contamination by stars from the stellar disks and halos of the hosts, many of which are old RGB stars of the type that we hope to use to detect dwarf galaxies. Since we want to look for overdensities relative to the local background, we apply a significance mapping technique to account for this spatially-varying interference. \par

To estimate the significance of features in our smoothed stellar density map, we apply the technique of \cite{Walsh2009}, which estimates the significance, $S$, of pixel \emph{i, j} of the map $M$ as the number of standard deviations above the local mean,
\begin{equation} \label{equation:walsh}
  S_{i,j} = \frac{ M_{i,j} - \overline{M}}{M_{\sigma}},
\end{equation}
\noindent where $\overline{M}$ and $M_{\sigma}$ are the local mean and standard deviation of the stellar density map around pixel $i, j$, calculated using pixels within twenty times the smoothing kernel bandwidth. This is a smaller background region than many large-area surveys use, but our background varies on a sufficiently small scale that this value was found to be optimal through the image simulations discussed in \S \ref{Section: selection}. \par

After creating these significance maps, we create segmentation maps of contiguous regions with $S_{i,j} > 3.0$, from which we derive our candidate list. We set a minimum size for acceptance of 10 contiguous map pixels (typically $\sim29$ arcsec$^2$), which rejects occasional catalog artifacts like bright stars or background galaxies shredded into multiple sources by \textsc{daophot}. This lower limit to detection size corresponds to a half-light radius of 55 pc or $3\arcsec$ at a distance of 3.8 Mpc, well below the typical size for dwarf galaxies and the size at which we expect to be able to resolve individual stars in the dwarfs. We therefore expect this limit to have a negligible effect on our ability to detect dwarfs, while allowing us to limit false detections in our candidate catalogs.

\section{Dwarf Completeness Tests} \label{Section: selection}
For comparisons with theoretical SLFs, we need to understand our observational completeness function (or selection function) as a function of dwarf properties. In particular, the size, luminosity, star formation history (SFH), and spatial location of dwarfs are the properties which are most likely to impact our completeness. Dwarfs with half-light radii within an order of magnitude of the seeing will have most of their stars blended, making it difficult to detect individual stars, while dwarfs that are large and faint may have their signal washed out by the high density of background sources. Meanwhile, dwarfs with SFHs very different from our search isochrones may have their stars underweighted by the matched-filter algorithm, masking their signal. Finally, due to our relatively small field of view and the large angular area subtended by our hosts, the background source density and its color-magnitude diagram (CMD) vary significantly as a function of spatial position, meaning an identical dwarf will be harder to detect if it is located near the host disk than if it is near the edge of our field of view. \par

We use two methods to estimate our completeness -- image simulations and catalog simulations. Image simulations are the traditional approach, and they allow the most accurate measurement of detection efficiency since dwarf models are recovered from the image simulations in an identical fashion to how real candidates would be detected. However, they are computationally expensive and a full exploration of the parameter space for dwarf recovery may be prohibitive for deep, wide surveys. Catalog simulations offer an alternative, in which all analysis is performed on the catalog level and observational effects like photometric completeness, error, and blending must be modelled explicitly. In return for their modelling complexity, catalog simulations are much faster, allowing a more complete sampling of the parameter space of dwarf properties. Models of the observational errors, completeness functions, and blending effects are used to generate a catalog that approximates what would be recovered by the photometric pipeline. The dwarf detection algorithm is then applied to this processed catalog to determine whether the simulated galaxy is recovered or not. We discuss the methods for our image simulations and the catalog simulations separately, then compare their performance and results. 

\subsection{Image Simulations} \label{subsection:image_simulations}
The image simulations are built on the same principles as those presented in \S 3.1 of \cite{Garling2020} -- mock galaxies are generated star by star, with positions sampled from Plummer profiles \citep{Plummer1911,Dejonghe1987} and magnitudes sampled from the \cite{Chabrier2001} lognormal initial mass function with PARSEC 1.2S isochrones \citep{Aringer2009,Bressan2012,Chen2014}, including the thermally-pulsating asymptotic giant branch and other improvements from \cite{Marigo2017}. We adopt the same fiducial 10 Gyr, [Fe/H] = $-2$ isochrone as used in our dwarf search for the bulk of the image simulations. Stars are sampled from the isochrone iteratively until a predetermined $M_V$ is reached. The full range of parameter space we explore is given in \S \ref{subsection:completeness_comparison}. We refer to this initial stellar catalog, before any manipulation, as the ``pure'' catalog. Galactic reddening is applied and these mock galaxies are then injected into the science frames using \textsc{addstar}, as in the artificial star tests described in \S \ref{subsection:artificialstartests}. We choose random positions for the dwarf galaxies to sample our spatial completeness. Stellar magnitudes and positions are shifted to the adopted distances of the hosts.\par

\subsection{Catalog Simulations}
For the catalog simulations we build pure dwarf star catalogs as described above. We then use probabilistic models to simulate the effects of blending, point-source completeness, and photometric error. We first simulate the effect of unresolved blending on the full, pure catalog. Then we sample the blended catalog randomly using the completeness functions from the artificial star tests modulated by an adjustment based on the local surface brightness. When the local surface brightness is high (as it is for compact dwarf galaxies), the completeness is reduced, accounting for the dependence of \textsc{daophot}'s detection algorithm on the local surface brightness. We add photometric errors to the blended, completeness-sampled catalog by sampling from the artificial star tests at the magnitudes of the catalog stars to get the final stellar catalog. This is embedded in a background sampled from the science images, and our full detection pipeline is run on this simulated catalog to determine if the dwarf is detected or not.\par

Our catalog simulation is similar in spirit to that of \cite{DrlicaWagner2020}, although we we developed our method independently. The most important aspect where our approaches differ is in the treatment of blending. \cite{DrlicaWagner2020} searched for satellite galaxies of the Milky Way, and so could safely neglect blending for typical dwarf sizes. As we are searching for dwarf satellites of hosts at distances of 3--4.4 Mpc, a classical dwarf galaxy with a half-light radius of 400 pc spans only $\sim20^{\prime\prime}$. In our ground-based imaging classical dwarfs will not be fully resolved, and anything more compact will be heavily blended. This can affect the color, TRGB magnitude, and stellar luminosity function of the red giant branch. Thus, we must consider the effects of blending in order for the catalog modelling approach to be robust. \par

The advantage of this technique is speed; image simulations are slow mainly because they require running photometry on each simulated image, and photometry is one of the slowest steps in our image processing pipeline. Additionally, very few stars in the image will actually be stars from the injected dwarf -- thus much time is spent inefficiently re-measuring field sources. An unoptimized implementation of our algorithm in the Python language shows a 100x speed-up for faint dwarfs ($\text{M}_{\text{V}}\approx-6$), and a more moderate speedup of about 20x for brighter dwarfs ($\text{M}_{\text{V}}\approx-10$). It is worth noting that the runtime of image simulations scales weakly with dwarf luminosity as it is nearly always the case that field sources dominate the runtime. However, as our catalog method includes blending, it exhibits fairly strong scaling with dwarf luminosity due to the necessity of computing the projected separations of stars. A detailed efficiency analysis is beyond the scope of this work, but may be explored in future work.\par

\subsubsection{Blending}
For our purposes we consider two types of blending; resolved blending and unresolved blending. We consider resolved blending to occur when two stars are detected and measured by the photometric pipeline, but where their PSFs overlap and light from the stars can be confused. \textsc{daophot} has excellent crowded-field performance, at least in part due to the fact that it measures stars simultaneously in local groups to account for sources with overlapping PSFs. We will assume \textsc{daophot} treats these situations properly and do not explicitly model this contribution -- this is motivated by the fact that unresolved blending presents a much bigger problem.\par

We consider unresolved blending to arise when a source is detected and measured by the photometric pipeline while there is another, undetected, source contributing flux to pixels within that star's PSF. This will erroneously make the detected source brighter in each band, but if the colors of the two stars are sufficiently different it can also change the apparent color of the detected source. If we again consider our $r_h=400$ pc $=20^{\prime\prime}$ classical dwarf, we could only fit $\sim330$ stars into that angular area if they are separated by one full width at half maximum (FWHM). However, assuming a stellar mass of $10^6$ M\textsubscript{$\odot$}, comparable to Leo II or Draco \citep{Woo2008}, there should be more than $10^3$ stars brighter than an apparent magnitude of 27 (about our 50\% completeness limit) at 3.5 Mpc and more than $10^6$ stars all the way down the main sequence for an old stellar population (for an observational comparison, see Figure 8 in \citealt{Dolphin2002a}; they have more than 5000 stars above the main sequence turn-off for Leo II). Thus we expect the bright RGB stars we seek to detect to have their measured magnitudes affected by many unresolved sources. We seek to understand how this affects our detection efficiency. \par

\begin{figure}
	\centering
        \includegraphics[width=0.45\textwidth,page=1]{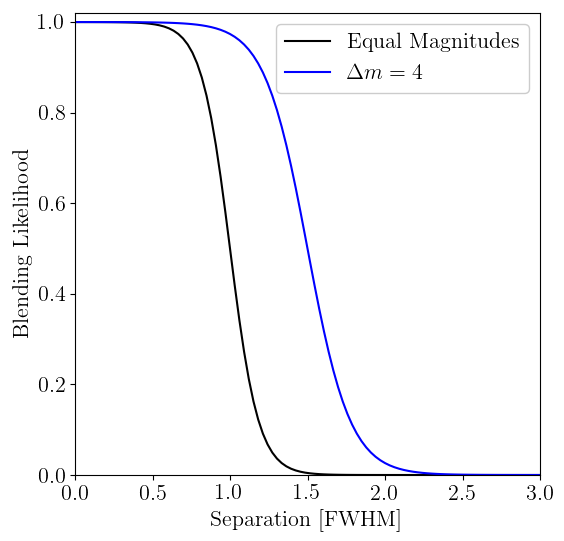} 
	\caption{Estimates of the likelihood for two stars to exhibit an unresolved blend ($\mathcal{L}$ in Equation \ref{equation:blending_likelihood}) in our data, at two extremes: when the two stars being considered are the same magnitude (black) and when they have a difference of 4 magnitudes.} 
	\label{figure:blending_likelihood}
\end{figure}

\begin{figure*}
	\centering
        \includegraphics[width=0.45\textwidth,page=1]{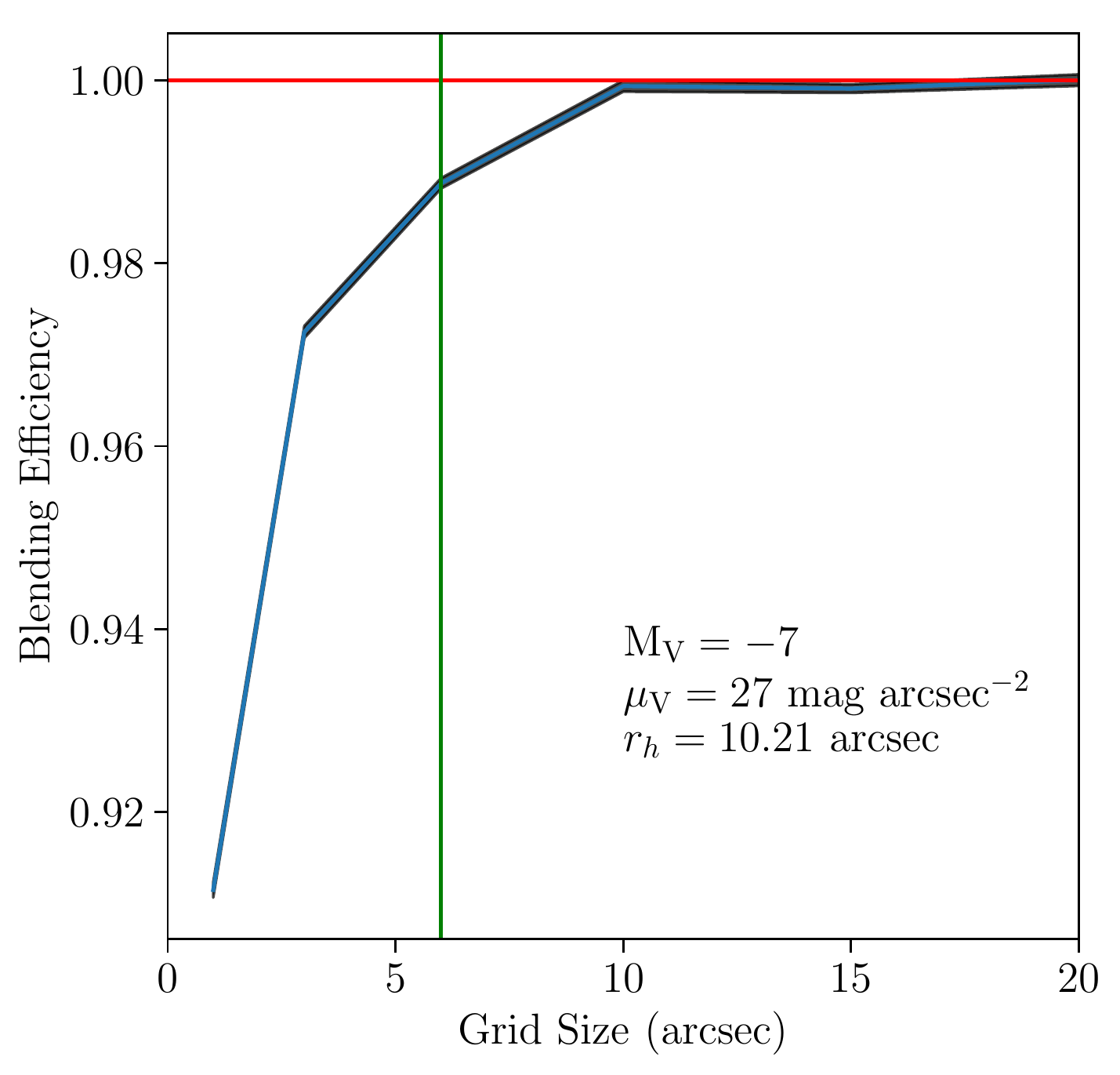} 
        \includegraphics[width=0.45\textwidth,page=2]{figures/blending_grid_experiment_minus7_8.pdf} 
	\caption{ The efficiency of our blending algorithm as a function of the fixed grid size for two dwarf models at 3.04 Mpc. The left model has $M_{\text{V}}=-7$ while the right model has $M_{\text{V}}=-8$, and both models have $\mu_{\text{V}}=27$ mag arcsec$^{-2}$. We define the blending efficiency to be the ratio between the number of blended stars at a particular grid size to the number of blended stars for an ungridded calculation (i.e., grid size $\geq$ galaxy size). We capture 98-99\% of all blends at our fiducial grid size of $6\farcs6$. The black shaded region shows the 1-$\sigma$ uncertainties derived from 100 iterations performed at each grid size; run-to-run variations in blending efficiency at fixed grid size are extremely small. }
	\label{figure:blended_test}
\end{figure*}

The likelihood of unresolved blending should depend, at minimum, on the angular distance between two sources and the seeing of the image. We also include a dependence on the magnitude difference between the stars, as it is more difficult to detect a faint star next to a bright one at fixed angular separation compared to two stars of comparable brightness. Generally there should also be a dependence on the pixel scale -- at a fixed angular FWHM, a smaller pixel scale may allow the detection algorithm to separate an unresolved blend into a resolved blend. Our prescription for the likelihood of blending two stars is calibrated roughly on idealized LBT/LBC image simulations which we analyzed with \textsc{daophot}. We define the probability that two stars of magnitudes $m_1$ and $m_2$ with magnitude difference $\Delta m=|m_1 - m_2|$ and an angular separation of $\theta$ will be blended as

\begin{equation} \label{equation:blending_likelihood}
  \begin{aligned}
    \mathcal{L}(\theta) &= \dfrac{\displaystyle 1 + \frac{\exp{\left(-m_{50}\right)}}{\rho}}{\displaystyle 1 + \frac{\exp{\left(\theta-m_{50}\right)}}{\rho}} \\
    a(\Delta m) &= 
    \begin{cases}
      1 \ ,& \text{if } \ \Delta m \leq0.5 \\
      0.5 \, \Delta m + 0.75 \ ,& \text{if } \ 0.5 < \Delta m \leq 1.5 \\
      1.5 \ ,& \text{if } \ 1.5 < \Delta m \\
    \end{cases} \\
    m_{50} &= a \times \text{FWHM} \\
    \rho &= a \times P / \text{FWHM} \\
  \end{aligned}
\end{equation}

\noindent where FWHM is the angular FWHM of the image PSF and $P$ is the pixel scale of the image. The factor $a$ controls the separation at which there is a $50\%$ probability for two sources to be blended and modifies the steepness of the likelihood function. $\mathcal{L}$ becomes a step function for small values of FWHM, $P$, and $\theta$, so we scale these variables together until FWHM $\geq2$ to produce a smooth curve. An estimate of $\mathcal{L}$ for our data is shown in Figure \ref{figure:blending_likelihood}. \par

We now describe our algorithm for blending the pure stellar catalogs. The only inputs not present in the pure catalogs are the pointwise separations $\theta$ between stars required for Equation \ref{equation:blending_likelihood}. For computational efficiency, the simulated CCD chips are gridded into segments of height and width equal to six times the imaging FWHM ($6\farcs6$). Blending is computed independently for each segment in the fixed grid, reducing the number of pointwise distance calculations needed for the evaluation of the blending likelihoods. In each grid segment, pointwise distances between all catalog sources are calculated and blending is evaluated in order of ascending brightness (or, descending magnitude). The blending is assigned probabilistically according to the blending likelihood function. The new, blended source is placed at the same location as the brightest star involved in the blend (the primary source), and so supersedes this source, while the fainter star is removed from the catalog. This blended source may then be blended with other sources. This procedure is repeated for each grid segment until the final, blended catalog has been produced. \par

\begin{figure*}
	\centering
        \includegraphics[width=0.45\textwidth,page=1]{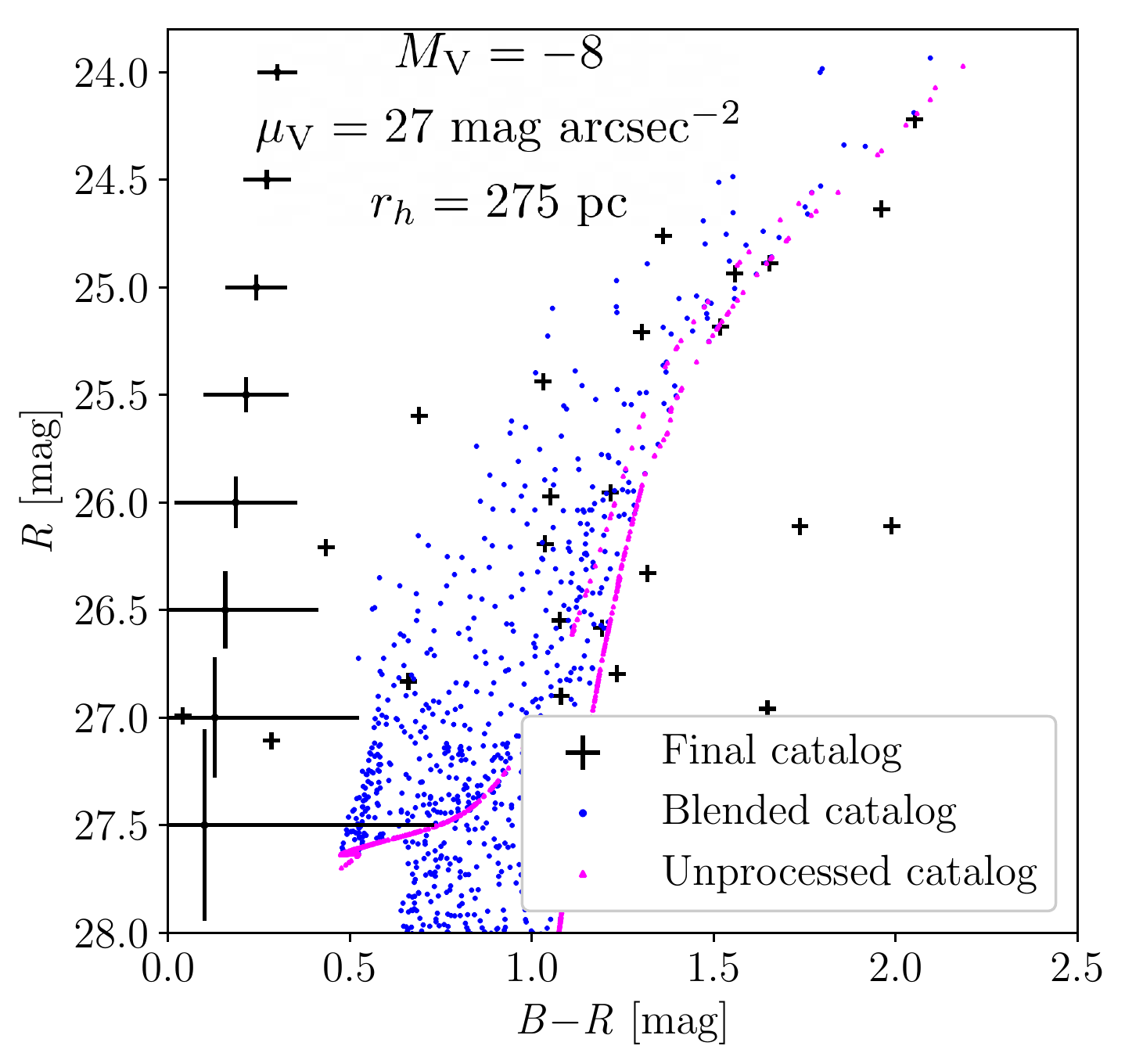} 
        \includegraphics[width=0.45\textwidth,page=1]{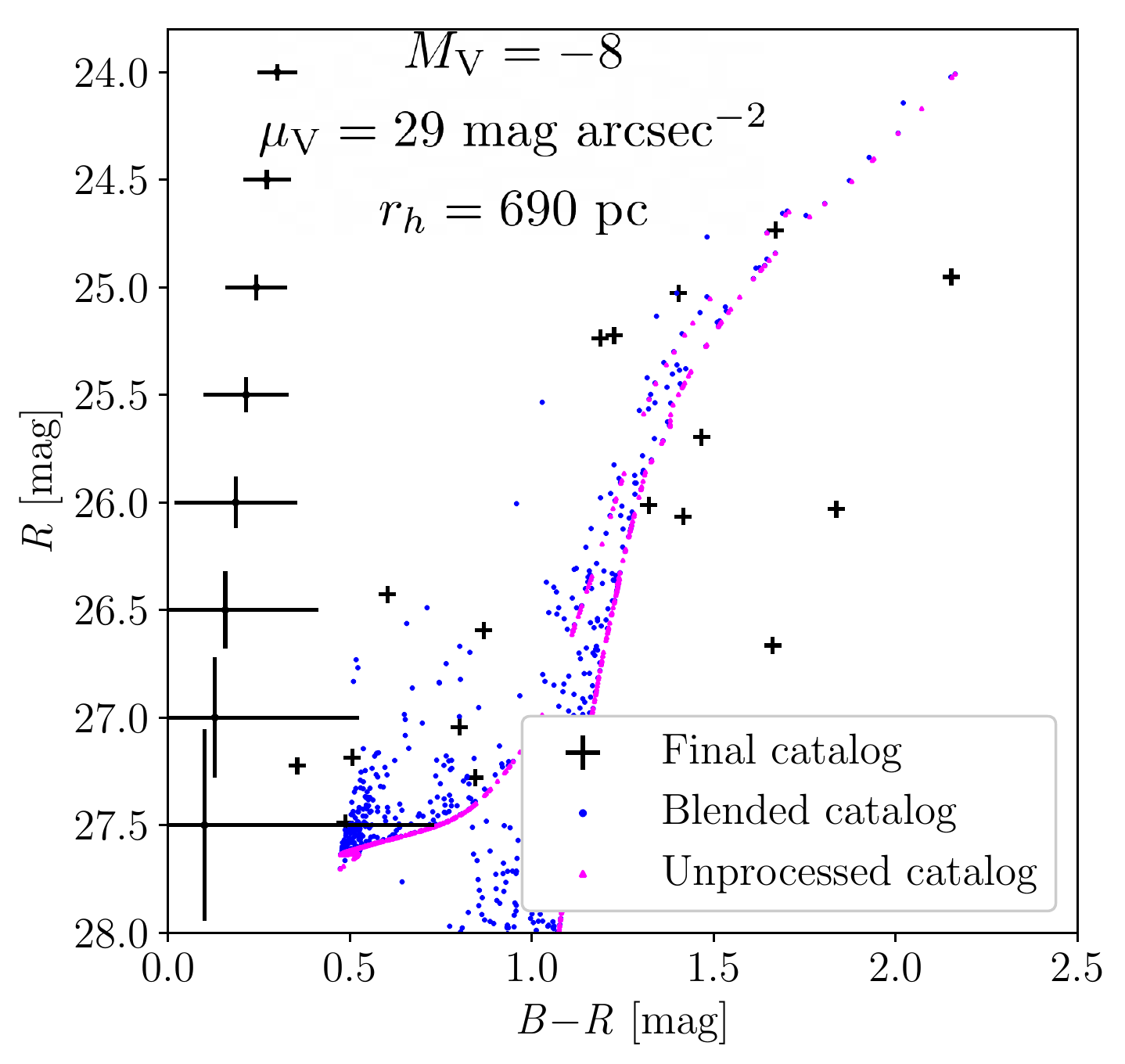} 
	\caption{Comparison of different stages of the catalog simulation process for $M_{\text{V}}=-8$ dwarfs of two different sizes at the distance of NGC 4214 (3.04 Mpc; our closest host). Magenta points show the pure dwarf stellar catalog before any processing. Blue points show the stellar catalog after blending but before application of the completeness and photometric error models. Small black crosses show the final photometric catalog after all these steps. Median photometric errors are shown along the left side of each plot. \emph{Left:} $r_h=275$ pc ($\mu_V=27$ mag arcsec\textsuperscript{-2}; a fairly typical size for dSphs of this luminosity). Stars around the MSTO contribute most to blending, as they are numerous and bright enough to affect the magnitudes of RGB stars. The color of the MSTO lies between $0.6\leq$ \emph{B}$-$\emph{R} $\leq1$ for this model stellar population, so RGB stars are shifted to bluer colors while the blue horizontal branch shows a slight redward shift. \emph{Right:} $r_h=690$ pc ($\mu_V=29$ mag arcsec\textsuperscript{-2}; a bit large for this luminosity). With a larger $r_h$, we see much less blending; as a result, there are fewer stars in our observable magnitude range because fewer stars are blended up to brighter magnitudes, but the detected stars are generally closer to the actual RGB. Some stars are scattered redward in the final catalog due to the addition of photometric errors.}
	\label{figure:blended_catalog}
\end{figure*}

Use of a fixed grid can give rise to resolution effects; in this case, stars that have small separations but lie on opposite sides of a grid line cannot be blended. Thus, use of a grid will mean stars along the grid lines may be under-blended, leading to a final blending catalog in which the number of blended stars depends on the grid size chosen. However, as the grid size decreases, fewer pointwise distances need to be computed for each grid segment, lowering per-process memory usage, and fewer pointwise distances need to be computed overall, decreasing runtime. We therefore need to strike a balance between accuracy and computational efficiency. \par

In Figure \ref{figure:blended_test} we show the dependence of the blending efficency on the size of the blending grid segements, where the blending efficiency is the ratio of the number of blended stars at a particular grid size to the number of blended stars for a fully ungridded (i.e., grid size $\geq$ galaxy size) calculation. We show two model galaxies, with M$_{\text{V}}=-7, \ -8$ where both have $\mu_{\text{V}}=27$ mag arcsec$^{-2}$. For these models, we are capturing 98-99\% of the blends at our fiducal grid size of $6\farcs6$. Calculations for other models show that the higher the average stellar density within the dwarf (i.e., lower M$_{\text{V}}$ or $\mu_{\text{V}}$), the more deficient the blending at fixed grid size. \par

We show realizations of our catalog simulation process with the fiducial grid size in Figure \ref{figure:blended_catalog} for dwarfs with identical absolute magnitudes but different sizes, illustrating how the surface brightness of the dwarf affects the blending calculation. As expected, we see that the more compact dwarf exhibits more unresolved photometric blending. Interestingly, the additional flux added to faint stars by unresolved blending in the compact model causes there to be more bright sources overall in the part of the CMD that we are sensitive to. However, this is mitigated by the surface-brightness-dependent completeness correction we develop in the next section, such that similar numbers of sources are detected in both models. Thus the most significant difference between these models is the shift of the RGB in the blended catalog to bluer colors for the compact dwarf model due to the flux contributed from stars around the main sequence turn-off. Reanalyzing the same dwarf models used to create Figure \ref{figure:blended_catalog} with a larger grid size of $40^{\prime\prime}$ showed no noticable changes in the blended CMD or the detection statistics, indicating that our fiducial grid size of $6\farcs6$ is sufficient to resolve the blending behavior.\par

The problem of crowding has been studied statistically by \cite{Olsen2003} and references therein, which give analytic formulas for the photometric errors due to blending. Given how faint our target stars are ($24<m<27$ mag), our measurement errors are comparable to the expected errors due to crowding for typical dwarf surface brightnesses. For example, at one magnitude below the TRGB ($m_{\text{V}}=25.7$ mag), our typical photometric errors from the uncrowded artificial star tests are $\sigma_{\text{V}}=0.2$ mag. For a star of the same magnitude, an equivalent error due to blending \citep[Equation 8 in][evaluated with the luminosity function from our model isochrone]{Olsen2003} is found at a surface brightness of $27.5$ mag arcsec$^{-2}$, typical of the dwarfs we hope to detect. Due to the width of the matched filter used in our detection algorithm scaling with the errors in the artificial star tests, it is possible that blended stars may be scattered away from the filter by this additional source of error. However, as shown in Figure \ref{figure:blended_catalog}, many of these blended sources still yield measurements near the model isochrone, and can therefore contribute signal to the detection algorithm. \par

Many other features of the blended CMDs in Figure \ref{figure:blended_catalog} resemble those found in earlier works \citep[e.g.,][]{Olsen2003,Gallart1996}. Inherently blue stars (i.e., stars on the blue horizontal branch) are scattered redward while redder stars (i.e., stars on the RGB) are scattered to bluer colors due to flux contribution from stars on the upper main sequence. Intrinsically bright stars are affected less by blending than faint stars, shown by the similar TRGB magnitudes in the initial and blended catalogs and the smaller color dispersion at the bright end of the blended RGB. Overall, our blending method captures all the main effects in the CMD noted by previous works, but further examination of the method with respect to photometric blending errors (including individual filters and colors) may be necessary in the regime where the blending errors are expected to be much greater than the basic measurement errors, such as in the study of MW globular clusters with ground-based imaging.\par

\subsubsection{Photometric Completeness Effects} 
We now sample the blended catalog according to the completeness functions derived from the artificial star tests. However, we cannot adopt these directly -- the local surface brightness around detectable stars in our simulated dwarf galaxies will be elevated by light from unresolved stars. We refer to galaxies in this regime as being semi-resolved. Generally, dwarfs with half light radii less than 1 kpc will be semi-resolved in our imaging. However, the artificial star tests presented in \S \ref{subsection:artificialstartests} are designed to measure photometric performance for stars in non-crowded regions, as the artificial stars are placed far apart on a regular grid. Therefore, they do not take into account local surface brightness effects and instead give results that are representative for the average surface brightness of the image (typically, the sky brightness). \par

\begin{figure}
	\centering
        \includegraphics[width=0.4\textwidth,page=1]{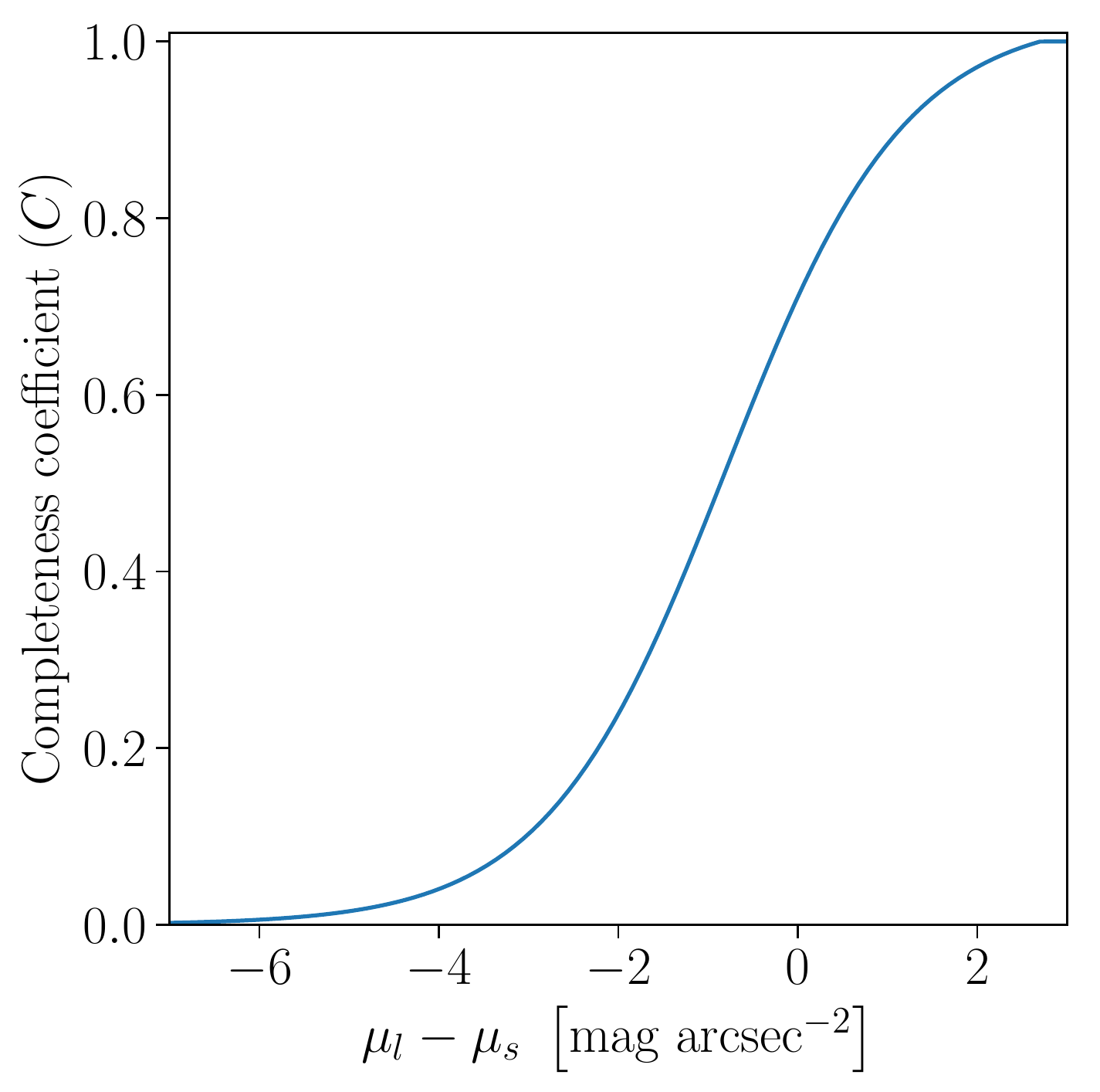} 
	\caption{Value of the multiplicative constant $C$ that modulates the completeness values from the artificial star tests as a function of the local catalog surface brightness ($\mu_l$) minus the average surface brightness of a star within the FWHM ($\mu_s$). This model lessens the likelihood of observing faint stars (high $\mu_s$) in regions of high local surface brightness (low $\mu_l$). This constant is calibrated to the image simulations and will likely need to be recalibrated for other datasets.} 
	\label{figure:completeness_coefficient}
\end{figure}

To account for this shortcoming, we implement a model that decreases the photometric completeness as a function of local surface brightness in the catalog. This model is tuned for our LBC data and calibrated from the image simulations in \S \ref{subsection:image_simulations}. We reduce the completeness from the artificial star tests by the multiplicative constant

\begin{equation} \label{equation:completeness_coefficient}
  \begin{aligned}
    & C(\mu_l, \mu_s) = \\
    & \begin{cases}
      \dfrac{\displaystyle 1 + \exp{ \left(-3.5\right)}}{\displaystyle 1 + \exp{ \left(\mu_s-0.8-\mu_l \right)}} \ ,& \text{if } \ \mu_l - \mu_s \leq 2.7 \ \text{mag arcsec}^{-2}\\
      1 \ ,& \text{if } \ \mu_l - \mu_s > 2.7 \ \text{mag arcsec}^{-2} \\
    \end{cases} \\
  \end{aligned}
\end{equation}

\noindent shown in Figure \ref{figure:completeness_coefficient}, where $\mu_l$ is the local surface brightness due to sources in the catalog and $\mu_s$ is the average surface brightness of a star within its FWHM, both in mag arcsec$^{-2}$.\par

We evaluate the local surface brightness in the catalog using the same grid as for the blending analysis to estimate the local surface brightness. Then we compute the values of our completeness functions from the artificial star tests at the magnitudes of each star in the catalog and multiply them by the $C$ factors for each band to find the final completeness estimates in each band. Three uniform variates are sampled for each star, and if two of them are less than the final completeness values for a star, that star is considered to be ``observed'' and will be included in the final catalog. This selection reflects the fact that we require detections in at least two bands to form our real photometric catalogs.\par

Finally, we add photometric errors to the stars that make it through the completeness analysis by performing a lookup against the artificial star tests. For each star, errors are sampled from artificial stars within $\pm \, 0.1$ magnitudes of the catalog values, maintaining the full error distribution. 

\begin{figure*}
	\centering
        \includegraphics[width=0.4\textwidth,page=2]{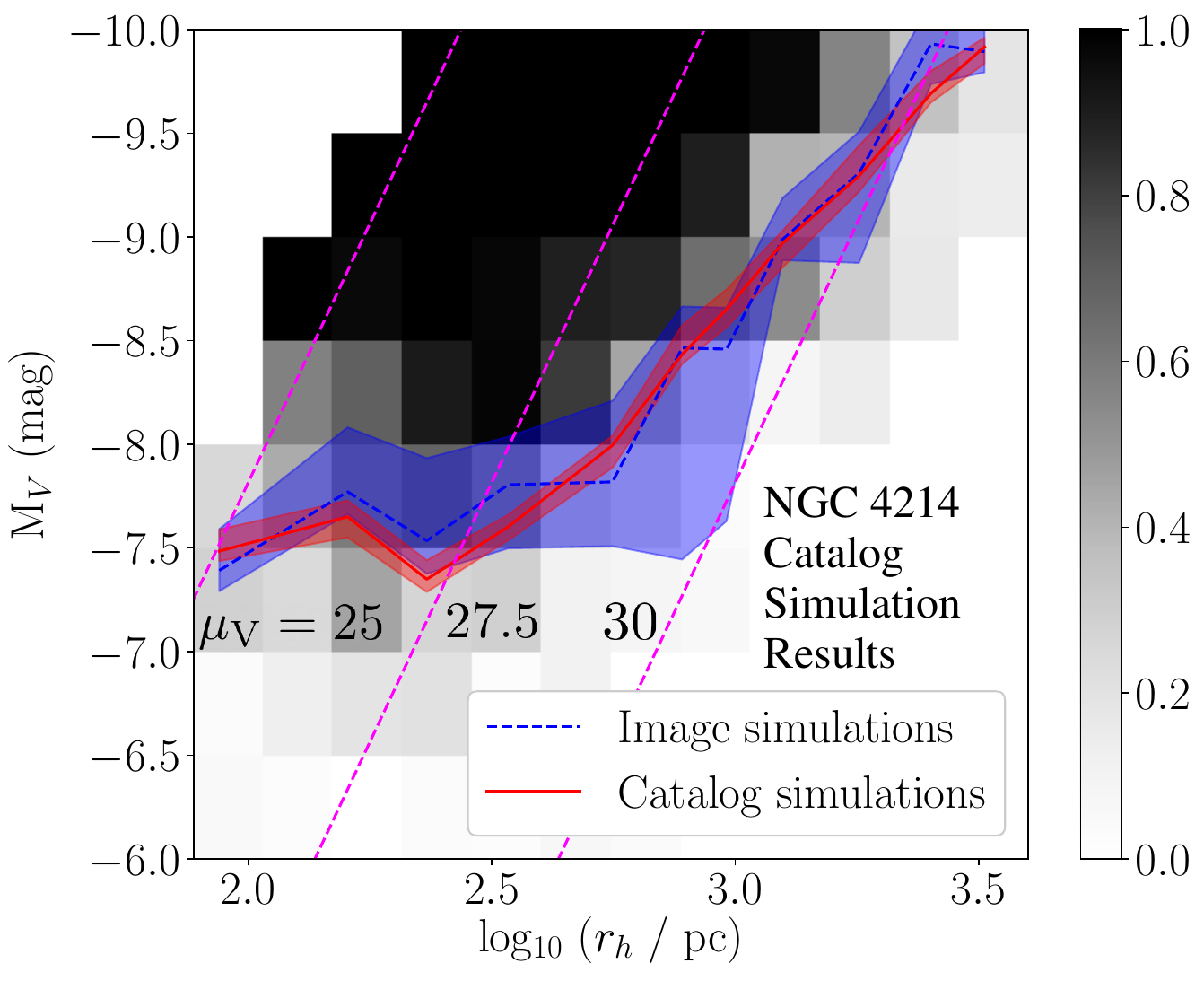} \hspace{0.5 in}
        \includegraphics[width=0.4\textwidth,page=2]{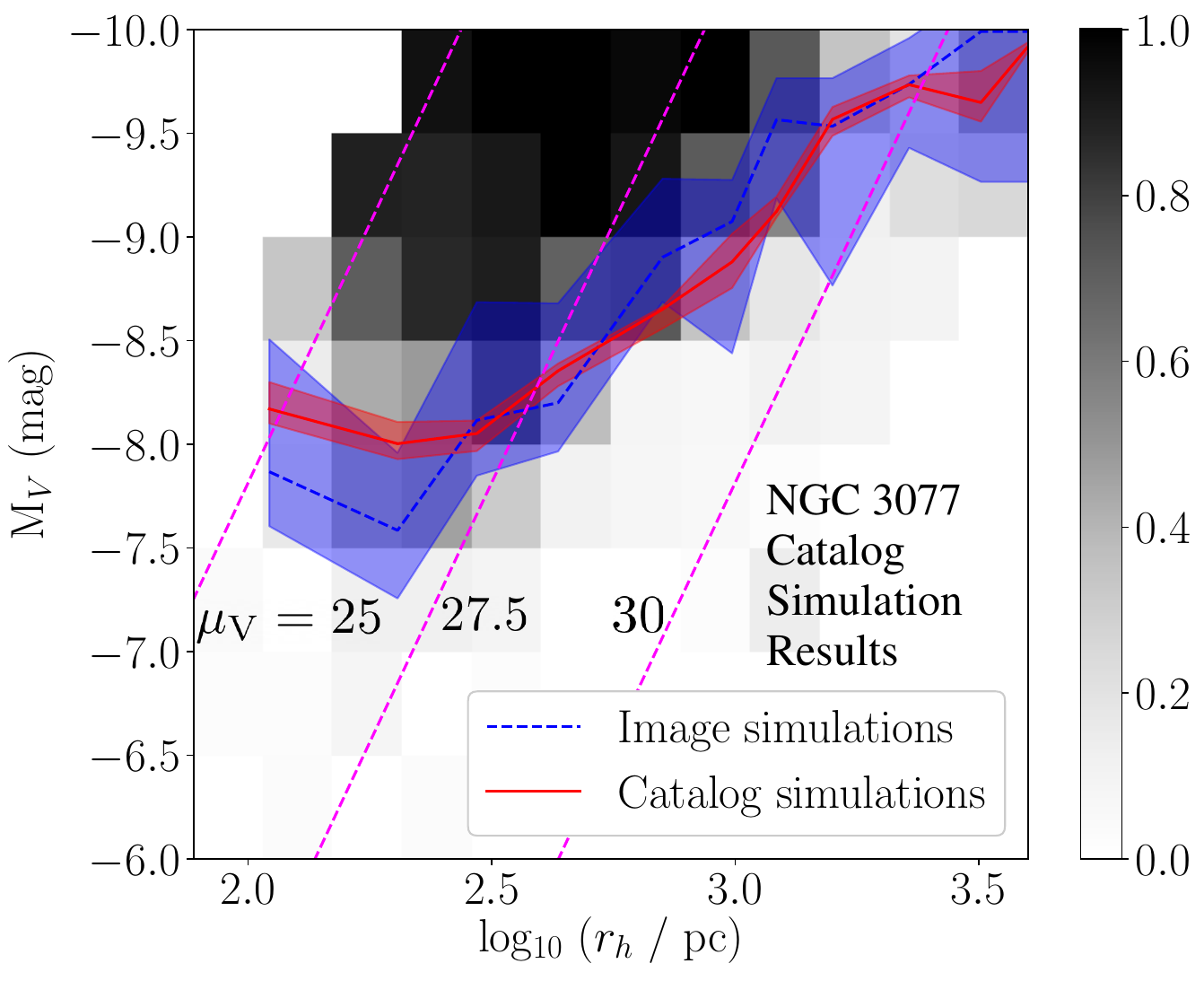} \\
        \includegraphics[width=0.4\textwidth,page=3]{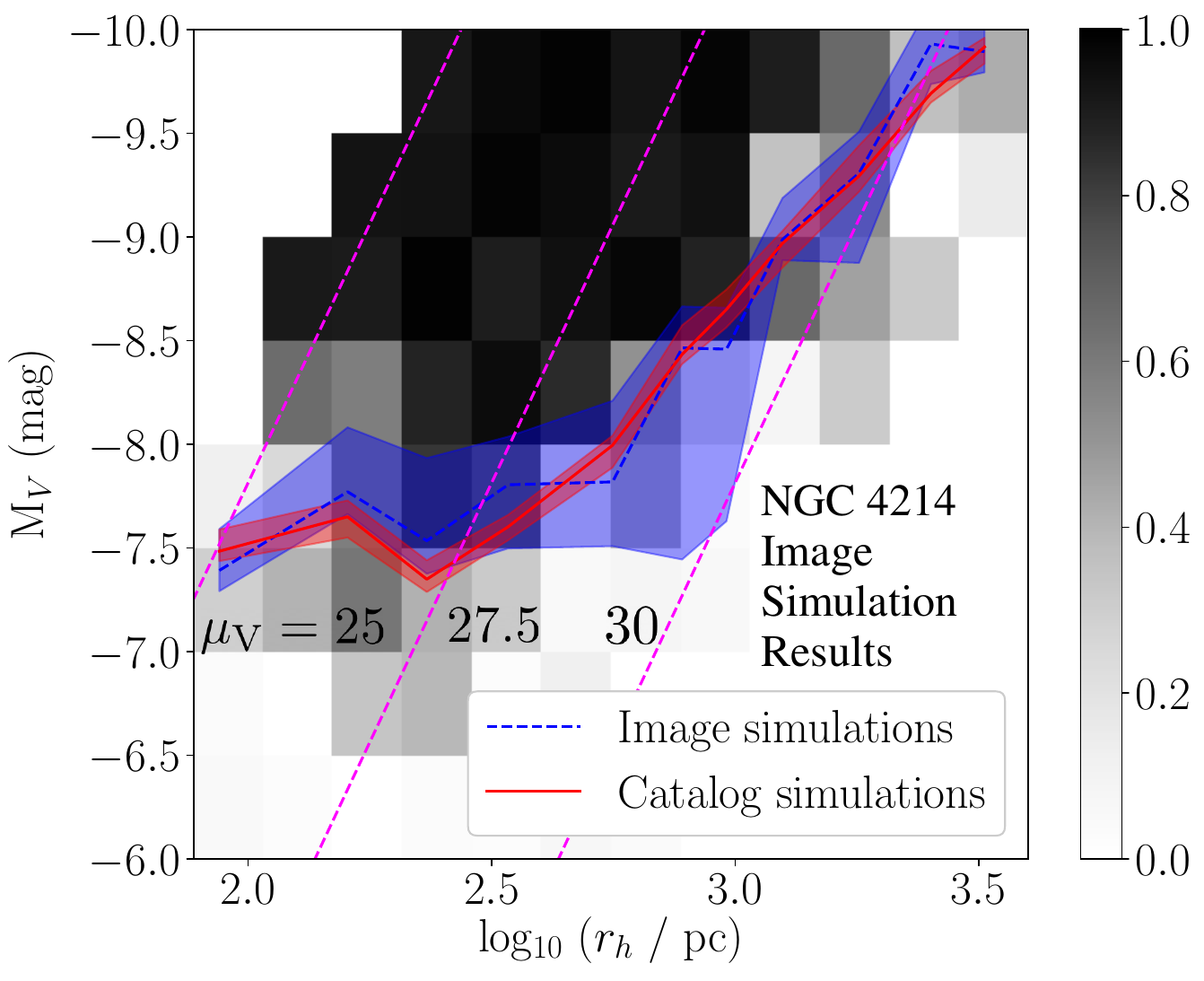} \hspace{0.5 in}
        \includegraphics[width=0.4\textwidth,page=3]{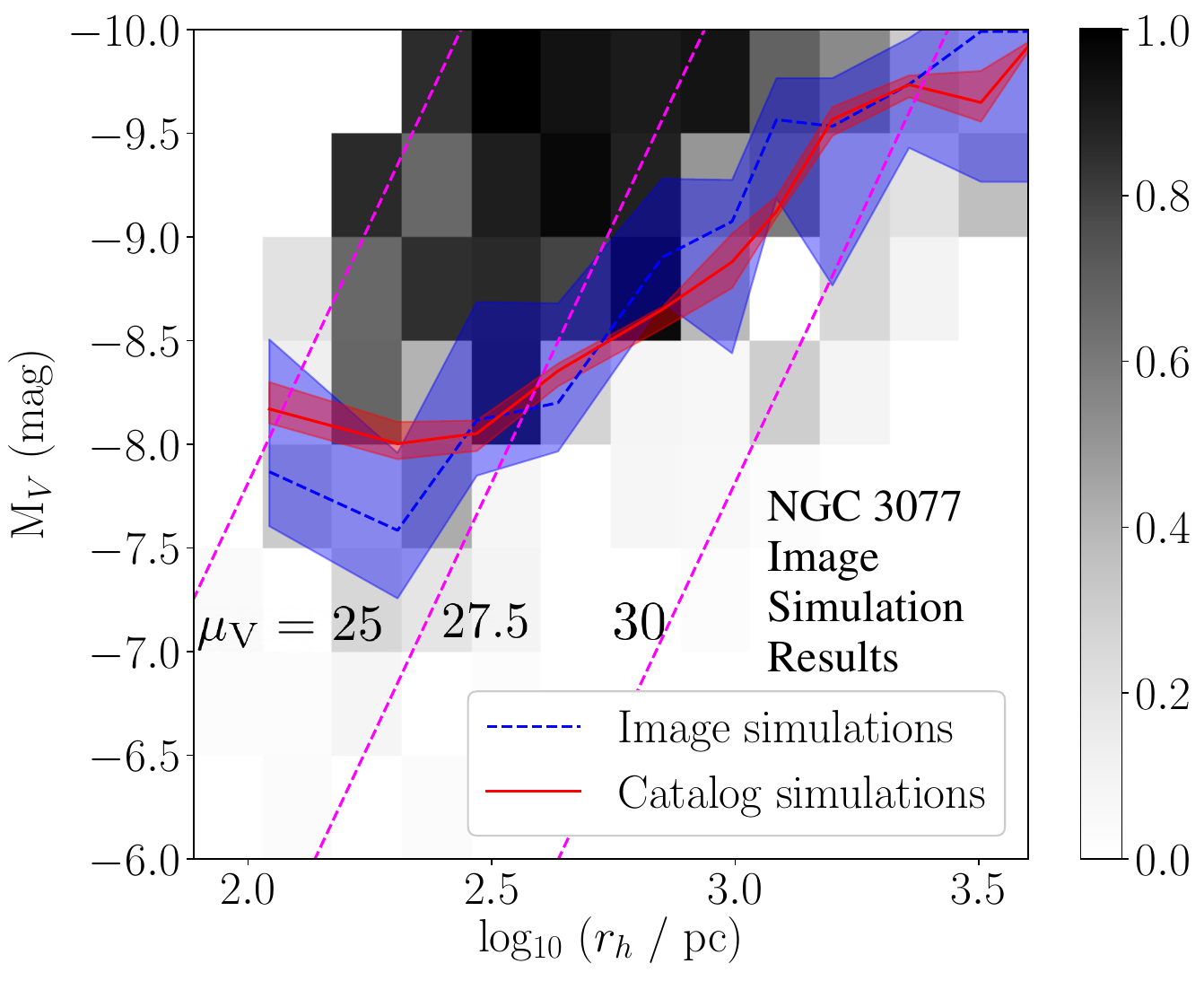} \\
	\caption{Results of catalog simulations (top row) and image simulations (bottom row) for NGC 4214 (left column) and NGC 3077 (right column). Approximate 50\% completeness curves are overlaid with statistical uncertainty ranges. Lines of constant surface brightnesses of 25, 27.5, and 30 mag arcsec\textsuperscript{-2} inside the half-light radius are shown in pink. Bins become darker following the colorbars as the completeness increases. The catalog simulations show more uniformity overall; they are nearly monotonically increasing to higher luminosities and smaller sizes. The image simulations show considerably more variation, which we attribute to variance in the detection efficiency with dwarf position. The catalog simulations use spatially uniform backgrounds for each realization, while the image simulations include the large local variations in the background due to contamination from stars belonging to the host.} 
	\label{figure:selection_comparison}
\end{figure*}

\subsection{Completeness Tests: Results and Comparisons} \label{subsection:completeness_comparison}
To examine our completeness as a function of the SFH, we ran a subset of image and catalog simulations using 1 Gyr, [Fe/H] = $-1$; 5 Gyr, [Fe/H] = $-1.5$; and 10 Gyr, [Fe/H] = $-2$ isochrones to generate the mock galaxies, as these bracket the range of average dwarf SFHs. We found that the significance values from Equation \ref{equation:walsh} were only affected at the $\sim 10\%$ level across this range of stellar populations. Given that the FWHM of the matched filter is typically about 0.2 mag at the TRGB, we conclude that the SFHs of the dwarf galaxies has little effect on our detection efficiency. We adopt the same 10 Gyr, [Fe/H] = $-2$ isochrone from the dwarf search method for the rest of our dwarf completeness tests.  \par

To sample our completeness in luminosity and size, we sampled galaxy luminosities uniformly from $-10 \leq \ M_{\text{V}} \  \leq -5$ mag and average surface brightnesses within the half-light radius log-uniformly from $25 \leq \mu_V \leq 31$ mag arcsec$^{-2}$. We ran 2000 image and catalog simulations each for NGC 4214 and NGC 3077 in order to have sufficient sampling in the full two-dimensional plane of $M_{\text{V}}$ and $r_h$ to show the differences between the methods. A comparison between the image and catalog simulations is shown in Figure \ref{figure:selection_comparison}. The catalog simulations show detection efficiencies which are generally smooth functions of $M_{\text{V}}$ and $r_h$, while the image simulations show significantly more scatter. We attribute this to the image simulations being embedded in the science images, where there is a complex spatially-varying background from the host. Due to this background, a galaxy that would nearly always be detected at the edge of our field may rarely be detected when it lies much closer to the host. While we sample the average background from the science images for our catalog simulations, it is spatially uniform for each simulation and so does not capture any spatial variation of the background source density. The detection efficiencies indicate our survey should be able to find analogs of dwarf galaxies like Ursa Minor, Draco, Carina, Tucana, and Leo T.\par

The most interesting features to compare between the methods are the gradient with respect to the surface brightness, the surface brightness of the faintest dwarfs that can be detected with $\geq80\%$ efficiency, and the change in the detection efficiency with respect to the half-light radius for small dwarfs. We see similar gradients with respect to the surface brightness between the catalog and image simulations, with both dropping fairly rapidly above $27.5$ mag arcsec$^{-2}$. Reproducing this trend required no particular calibration of the catalog modelling method, as the reduction in the detection efficiency is simply due to the decreasing signal-to-noise in the matched filter map. The catalog and image simulations for NGC 4214 show exactly the same combination of magnitude ($-8.5<M_{\text{V}}<-8$ mag) and size ($316<r_h<422$ pc) for the faintest dwarfs detected with $\geq80\%$ efficiency, while for NGC 3077 they differ by perhaps $0.5$ mag and $100$ pc. These matches are more difficult to reproduce, as the completeness limit depends on the density of the background and the treatment of blending. If we neglect blending entirely in the catalog modelling, the completeness limit moves to smaller sizes and fainter magnitudes, substantially biasing our estimates in this regime. Our treatment of blending is also critical for reproducing the decrease in the detection efficiency with respect to the half-light radius as we move to small sizes; this is the most difficult feature to match. Both the catalog-level blending and the surface-brightness-dependent completeness correction are needed to reproduce the decrease in detection efficiency measured in the image simulations, but they do surprisingly well. A particularly interesting result in the NGC 4214 simulations is that we simultaneously reproduce the steep drop in detection efficiency for dwarfs with $-8.5<M_{\text{V}}<-8$ inside of 316 pc and the much shallower drop for dwarfs with $-8<M_{\text{V}}<-7$, with the peak detection efficiency at fixed $M_{\text{V}}$ occurring at the same radius with both methods. With good agreement between the catalog and image simulations for all of these features, we use catalog simulations to derive the detection efficiencies for the rest of our hosts since they are much faster to compute compared to image simulations. Improving the background modelling is one of the main areas where we hope to improve the catalog simulations in the future.\par

\section{Results} \label{section:results}
\subsection{Observational Results}
\begin{figure}
	\centering
        \includegraphics[width=0.45\textwidth]{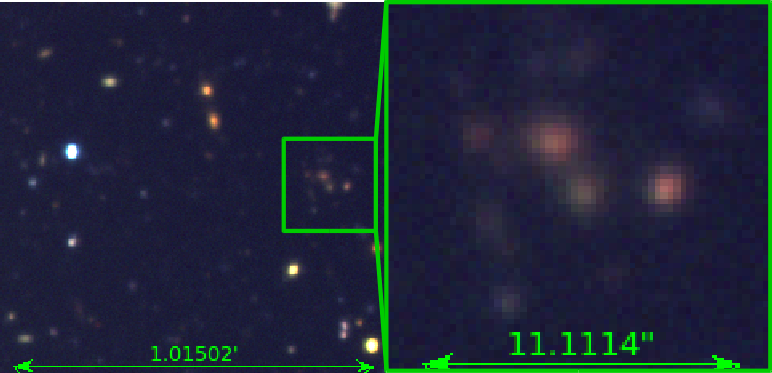} 
	\caption{Example of a false positive. The left panel shows a $\sim1^{\prime}$ field of view to illustrate the typical point-source density in this region, and the detection is enclosed within a square. In the right panel we show a zoom-in on this detection region. The point-sources here have colors close to our model isochrone, and they are close together, so our algorithm reports this as a dwarf galaxy candidate. However, we would expect to see unresolved emission for a dwarf galaxy of this size, but we detect none, and the point sources show no coherent radial density structure, leading to its rejection. } 
	\label{figure:interloper}
\end{figure}

After candidates are identified using the procedures outlined in \S \ref{section:dwarfsearch}, we validate the detections with a combination of visual inspection and structural parameter estimation. The visual inspection checks that the candidate is not an obvious false detection, like a large background galaxy shredded into many point souces by \textsc{daophot}. If the detection does not appear to be due to an artifact, we attempt to measure its structure using an elliptical Plummer model with the maximum-likelihood approach presented in \cite{Martin2008}. We detect an average of four artifacts and three candidates per host, including two previously known dwarf galaxies, DDO 113, a satellite of NGC 4214, and LV J1228+4358, a satellite of NGC 4449. Most other candidates show a relatively flat spatial profile consistent with a chance arrangement of point sources matching the filtered isochrone. There are a few candidates which show some spatial clustering of point sources that might indicate the prescence of a dwarf galaxy, but comparison with dwarfs of similar size from the image simulations described in \S \ref{subsection:image_simulations} reveals that we should expect detectable diffuse emission from these candidates, but we detect none. These candidates are most likely background galaxy clusters or groups of red galaxies. Analysis of the luminosity function of the point sources detected in these candidates also shows that there are more bright sources than we would expect at the distances of our hosts, further supporting this interpretation. An example of one such false detection is shown in Figure \ref{figure:interloper}.\par

\begin{figure}
  \centering
        \includegraphics[width=0.21\textwidth]{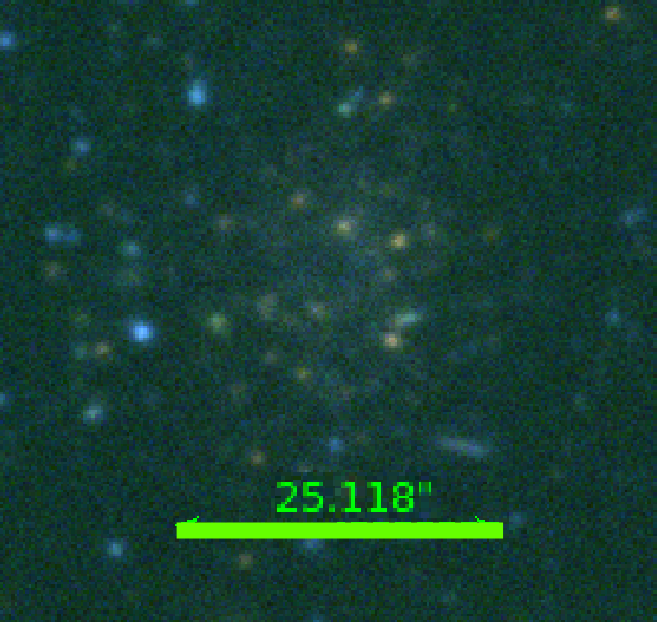} 
        \includegraphics[width=0.23\textwidth]{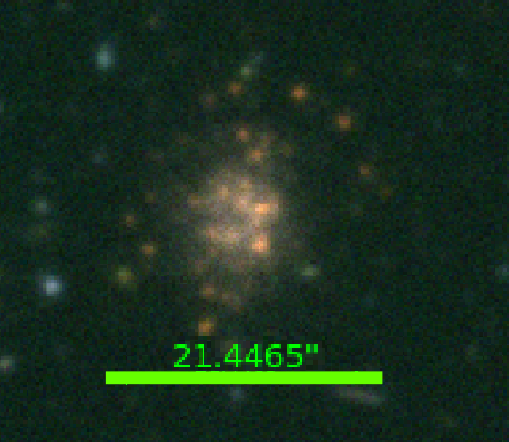} 
	\caption{Examples of the simulated dwarf galaxies used to derive our detection efficiency (second row of Figure \ref{figure:selection_comparison}) and to guide our visual inspections for NGC 4214. \emph{Left:} A $M_{\text{V}}=-8, r_h=280$ pc ($19^{\prime\prime}$) dwarf, for which we are $\sim50\%$ complete averaged over the full field. In regions of low source density (i.e., far from the host), we are significantly more complete, but such galaxies are rapidly lost near the host. There is some unresolved emission which can be visible depending on the local background level, and typically $10-20$ detectable RGB stars, although stochasticity in the IMF sampling affects our detection efficiency. \emph{Right:} A $M_{\text{V}}=-8, r_h=110$ pc dwarf ($7\farcs5$). Compared to the more diffuse dwarf, much more unresolved light is visible, but the center of the galaxy is badly blended and so the detection efficiency with the point-source algorithm is worse. We expect our visual inspection to help with the identification of these kinds of galaxies. 
}
	\label{figure:galaxy_examples}
\end{figure}

We also performed a full visual inspection of the images, focusing especially on small candidates, as our resolved-star search method loses some detection capability for dwarfs smaller than $\sim200$ pc (see Figure \ref{figure:selection_comparison}) due to the strong blending in these galaxies. To guide our visual search, we generated 42 image simulations for each host as in \S \ref{subsection:image_simulations} of dwarfs from $-10 \leq$ M$_{\text{V}} \leq -5$ mag and $25 \leq \mu_{\text{V}} \leq 31$ mag arcsec$^{-2}$. Examples of these model dwarf galaxies are shown in Figure \ref{figure:galaxy_examples}. These image simulations illustrate the expected visual appearance of any dwarfs, particularly how well-resolved they are at the distance of each host. We find an abundance of low surface brightness galaxies in our imaging, but they all have light profiles far smoother than we would expect if they were at distances of 3--4.5 Mpc. We identify no significant dwarf satellite candidates from this visual search. \par

We therefore consider only two satellite dwarf galaxies to have been found in our imaging survey, both of which were previously known: DDO 113, which is a satellite of NGC 4214 and was examined in \cite{Garling2020}, and LV J1228+4358, which is a large, disrupting \citep{Rich2012,Martinez-Delgado2012,Toloba2016a} satellite of NGC 4449. Both of these dwarfs are easily identified by our detection algorithm, with estimated detection efficiencies of $100\%$. Since LV J1228+4358 is in the process of tidal disruption, we must consider whether it should be characterized as a galaxy or a stream for the purposes of this analysis. Given that it appears to have a centrally-concentrated stellar component that may still be bound \citep{Martinez-Delgado2012}, we choose to treat it as a galaxy and include it in our theoretical analysis. It is worth noting that DDO 113 also shows evidence for interaction with its host, as its star formation was quenched about 1 Gyr ago \citep{Weisz2011}, likely because of cessation of cold gas inflows after its infall to NGC 4214 \citep{Garling2020}. Thus we may be seeing signatures that low-mass hosts affect their satellites more strongly than previously believed. \par

We adopt the measurements from \cite{Garling2020} for DDO 113, namely $m_{\text{V}}=15.17$ mag, so that its intrinsic luminosity is $M_{\text{V}}=-12.17$ at the TRGB-measured distance of 2.95 Mpc \citep{Dalcanton2009}. There is a measurement of $m_{\text{B}}=14.2$ for LV J1228+4358 in \cite{Karachentsev2013}, while \cite{Carlsten2019} measure $m_{\text{g}}=16.50$ and $m_{\text{i}}=15.95$, and \cite{Rich2012} and \cite{Martinez-Delgado2012} both find $M_r=-13.5$, $M_g=-13.05$. From our own imaging, we measure $m_{\text{B}}=15.69$ ($M_{\text{B}}=-12.35$) and $m_{\text{V}}=14.99$ ($M_{\text{V}}=-13.03$) by fitting S\`ersic profiles using \textsc{galfitm} \citep{Haeusler2013}, which is a multi-band extension of \textsc{galfit} \citep{Peng2010}; we perform this fitting analagously to the procedure described in \S3.1 of \cite{Garling2020}. While the \cite{Karachentsev2013} apparent magnitude is discrepant from our own, our photometry is consistent with the photometry of \cite{Rich2012}, \cite{Martinez-Delgado2012}, and \cite{Carlsten2019} given the \cite{Jordi2006} filter conversions. We adopt our measured quantities for LV J1228+4358 for the theoretical work in \S \ref{subsection:theory} Additional photometry and structural parameters are given in Table \ref{Table:photometry}. The structural parameters for LV J1228+4358 have large uncertainties due its irregular morphology.

\begin{table}
  \caption{Summary of \textsc{galfitm} photometry and S\`ersic structural parameters for detected dwarf satellites.}
  \scalebox{0.9}{
    \hspace*{-3.5em}
  \begin{tabular}{c c c c}
    \hline
    \hline
    Parameter & Value & Uncertainty & Reference \\
    \hline
    \multicolumn{4}{|c|}{DDO 113} \\
    \hline
    Host & NGC 4214 & & \\
    R.A. & $12^{\text{h}}14^{\text{m}}58.3^{\text{s}}$ & $0.1^{\text{s}}$ & \cite{Garling2020} \\
    Decl. & $+36\degree13^{\prime}07^{\prime\prime}$ & $2^{\prime\prime}$ & \cite{Garling2020} \\
    $m_{\text{U},0}$ & 15.62 & 0.10 & \cite{Garling2020} \\
    $m_{\text{B},0}$ & 15.66 & 0.10 & \cite{Garling2020} \\
    $m_{\text{V},0}$ & 15.16 & 0.10 & \cite{Garling2020} \\
    $m_{\text{R},0}$ & 15.09 & 0.10 & \cite{Garling2020} \\
    D & 2.95 \text{Mpc} & 0.083 \text{Mpc} & \cite{Dalcanton2009} \\
    $M_{\text{V}}$ & $-12.19$ & 0.18 &  \\
    $\text{R}_{\text{e}}$ & $43\farcs6$ & $1\farcs5$ & \cite{Garling2020} \\
    & 624 pc & 21 pc & \cite{Garling2020} \\
    n & 0.64 & 0.02 & \cite{Garling2020} \\
    q & 0.63 & 0.05 & \cite{Garling2020} \\

    \hline
    \multicolumn{4}{|c|}{LV J1228+4358} \\
    \hline
    Host & \text{NGC 4449} & & \\
    R.A. & $12^{\text{h}}28^{\text{m}}44.8^{\text{s}}$ & $0.1^{\text{s}}$ & This work \\
    Decl. & $+43\degree58^{\prime}06^{\prime\prime}$ & $4^{\prime\prime}$ & This work \\
    $m_{\text{B},0}$ & 15.69 & 0.20 & This work \\
    $m_{\text{V},0}$ & 14.99 & 0.20 & This work \\
    $m_{\text{R},0}$ & 15.11 & 0.20 & This work \\
    D & 4.07 \text{Mpc} & & \cite{Karachentsev2013} \\
    $M_{\text{V}}$ & $-13.03$ & 0.20 &  \\
    $\text{R}_{\text{e}}$ & $67^{\prime\prime}$ & $10^{\prime\prime}$ & This work \\
    & 1300 pc & 200 pc & This work \\
    n & 0.62 & 0.10 & This work \\
    q & 0.28 & 0.06 & This work \\    
    
    \hline
  \end{tabular} \\
  }
  \label{Table:photometry}
\end{table}

\subsection{Theoretical Results} \label{subsection:theory}
Here we present theoretical expectations for the cumulative SLF, both intrinsic and observed, of the LBT-SONG near-sample. We derive these expectations based on an expanded version of the Bayesian method presented in \cite{Dooley2017b}; this expanded formalism will be presented in an upcoming paper along with a theoretical interpretation of the full LBT-SONG sample. In brief, the method formulates a distribution function $d\text{N}/dM_{\text{V}}(M_{\text{V}}|\Theta)$ that can be integrated to yield the cumulative SLF, $N < M_{\text{V}} (M_{\text{V}}|\Theta)$, where $\Theta$ are model parameters. For this first analysis, we adopt a single set of model parameters, but we will explore other models in future work.

\begin{enumerate}
\item We start with the \cite{Moster2013} SMHM relation and assume a constant scatter in the relation of 0.2 dex \citep{Behroozi2013,GarrisonKimmel2017}, from which we form the probability distribution function (PDF) for the host halo mass given its observed stellar mass and uncertainty (see Table \ref{Table:observations}).
\item We adopt the \cite{Dooley2017b} subhalo mass function (SMF), which gives the distribution $d\text{N}/d\text{M}_{\text{sub}} \, (\text{M}_{\text{sub}}|\text{M}_{\text{host}})$ that describes the subhalo number distribution as a function of the host's halo mass. Marginalizing the product of the host halo mass PDF and the SMF over the host halo mass gives the mean SMF predicted for the host. We can transform this into observational units given the PDF for a subhalo of mass $\text{M}_{\text{sub}}$ to host a galaxy of luminosity $M_{\text{V}}$.
\item We adopt the reionization quenching model of \cite{Dooley2017b} to describe the likelihood that a subhalo of a given halo mass hosts a luminous stellar component.
  \item We then use the same SMHM relation and scatter described above to form subhalo stellar mass distributions given the halo masses.
  \item We next adopt the color-stellar-mass-to-light ratio relation with scatter from \cite{GarciaBenito2019} to model the PDF of the stellar-mass-to-light ratio as a function of color, and the empirical satellite color distribution from SAGA \citep{Mao2020}. Combining these allows us to construct a model PDF for the luminosity ($M_{\text{V}}$) of a satellite given its color and stellar mass. 
\end{enumerate}

\noindent Marginalization over the host halo and stellar masses, and the satellite halo masses, stellar masses, and colors yields $d\text{N}/dM_{\text{V}}(M_{\text{V}}|\Theta)$, which is then integrated to give the expected mean, intrinsic, cumulative SLF for each individual host. \par

\begin{figure}
	\centering
        \includegraphics[width=0.45\textwidth,page=1]{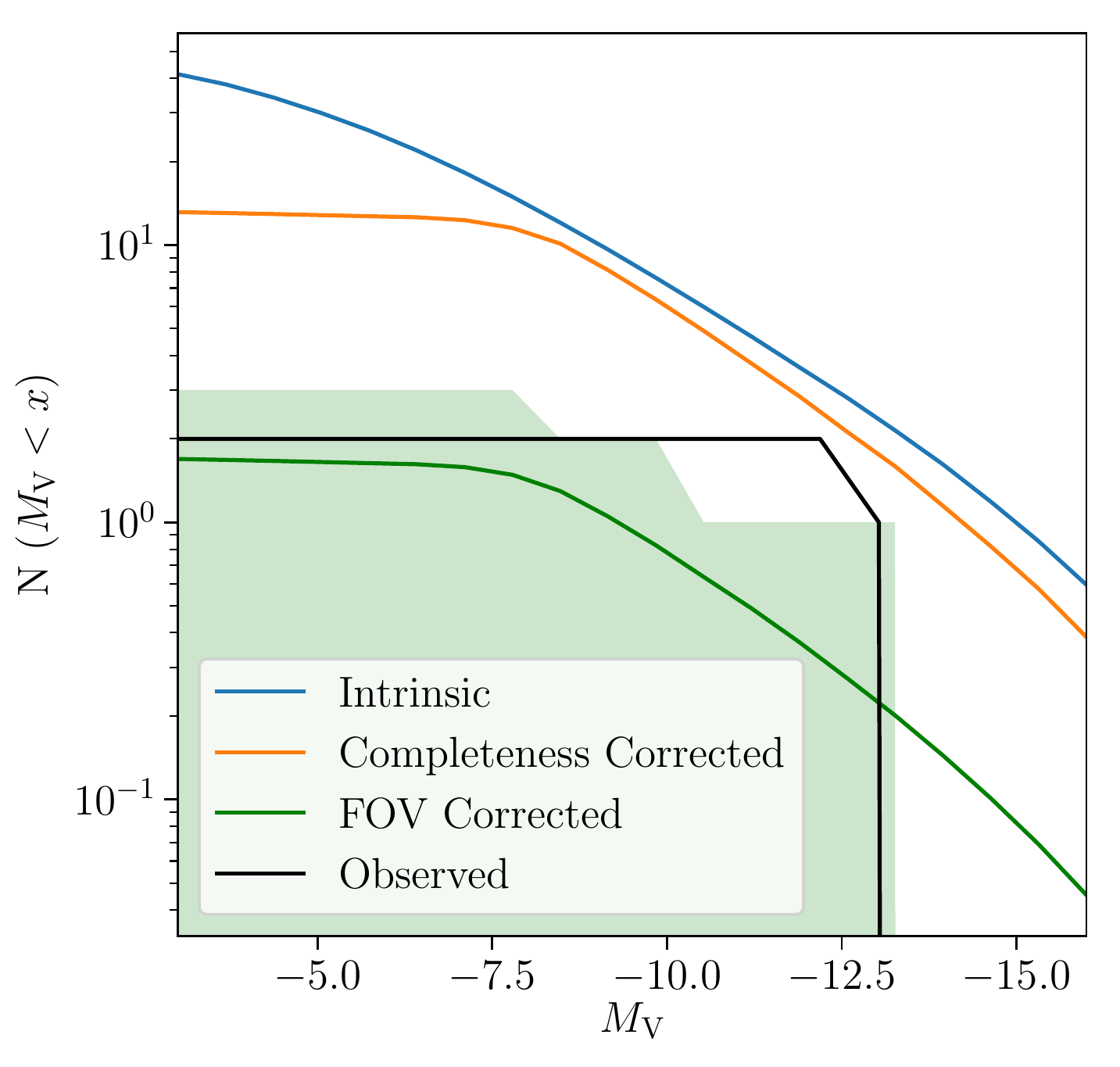} 
	\caption{Theoretical expectations for the cumulative SLF of the hosts in the LBT-SONG near-sample. We show the intrinsic SLF in blue, while the orange line shows the whole-halo expectation given our photometric completeness limits, and the green line gives the final expectation after factoring in our restricted field of view. The shaded region around the final expectation shows the 1-$\sigma$ range of SLFs from random samples given the expectation value. Our observed dwarf satellite sample, consisting of DDO 113 and LV J1228+4358, is shown in black.} 
	\label{figure:SLFs}
\end{figure}

We account for observational incompleteness in two steps; first accounting for photometric incompleteness within the field of view, then for the resticted field of view of our imaging. We can account for photometric incompleteness by using the fake galaxy tests conducted in \S \ref{Section: selection}, but we must include a treatment of dwarf galaxy size in the theoretical calculation to match the $r_h$ axis of our completeness tests. We estimate satellite galaxy sizes by adopting the simple model relating galaxy size to halo virial radius given by Equation 6 in \cite{Jiang2019}, $R_{e,3\text{D}}=0.02(c/10)^{-0.7} \, R_{vir}$, where $c$ is the halo concentration. However, converting $R_{e,3\text{D}}$ to 2D requires additional prefactors to account for projection effects ($f_p$) and a conversion from mass-weighting, which is used to derive $R_{e,3\text{D}}$ in simulations, to light-weighting, which is used observationally ($f_k$), such that $R_{e,2\text{D}}=f_k\,f_p\,R_{e,3\text{D}}$. We adopt $f_k\,f_p=0.78$ as in \cite{Somerville2018}, which is derived for predominantly spherical systems. \cite{Jiang2019} estimates a scatter in the galaxy-size-virial-radius relation of 0.18 dex, but we adopt 0.3 dex to account for random viewing angles and a range of possible galaxy shapes (e.g., additional spread in the $f_k\,f_p$ prefactor). We use the \cite{Child2018} concentration-mass relations, which are derived from dark-matter-only simulations assuming the WMAP-7 \citep{Komatsu2011} cosmology. During the integration, we parse the completeness tables and multiply our detection probability as a function of $M_{\text{V}}$ and $r_h$ into the joint PDF. \par

To account for our restricted field of view, we assume the radial distribution of satellites follows the mass distribution of the host halo. \cite{Han2016} showed that, when tracking all orphaned subhalos, the final radial distribution is more concentrated than the host halo. We assume some fraction of the orphaned satellites tracked into the inner halo in \cite{Han2016} and used to assemble their radial distribution will be disrupted and hence undetectable in real data. Additionally, our model is fairly similar to the observed radial distribution of satellites of the Milky Way (e.g., Figure 1 in \citealt{Kim2018a}; see also \citealt{Carlsten2020}). Correcting our distributions then amounts to calculating the fractions of the total host virial masses enclosed within our projected fields of view, and multiplying our distributions by these fractions. This treatment neglects possible splashback scenarios \citep[e.g.,][]{More2015,Diemer2020} by implicitly assuming that the virial radius is the halo boundary, but the internal consistency of the calculation is maintained as long as the same halo boundary is used for the calculation of the radial distribution and the SMF. Moreover, these corrections are small; including subhalos out to the splashback radius along the line of sight yields a $\sim5\%$ increase in expected satellite detections for a search over the full virial volume of the host, while the correction is negligible for our field of view \citep[Figure 11 in][]{Dooley2017b}. With this last correction, we produce the final estimates for the mean SLFs of the hosts in the LBT-SONG near-sample; the cumulative SLF for the entire host sample is shown in Figure \ref{figure:SLFs}. To estimate the intrinsic variance we should expect in our sample, we assume the SMF is negative binomial distributed as in \cite{BoylanKolchin2010}. We sample random SLFs from each host indepedently and sum them to form Monte Carlo realizations of the theoretical SLFs. The 1-$\sigma$ range of these samples is indicated by the shaded region in Figure \ref{figure:SLFs}. \par

The total number of dwarfs (2) is in good agreement between the data and the theoretical expectations, but the observed SLF has more classical dwarfs than expected, lying outside the 1-$\sigma$ range for dwarfs with $M_{\text{V}}<-10$. We find that only $4\%$ of our simulated samples contain two dwarfs brighter than $-12$ mag, as we find in our observational sample. Additionally, $55\%$ of simulated samples have at least one dwarf fainter than $-10$ mag while we find none in our observations. This disagreement in the shape of the SLFs is reflected in the results of the two-sample Kolmogorov-Smirnov test, which uses normalized cumulative distributions to estimate the probability that two samples are drawn from the same distribution. Comparing our Monte Carlo samples to our observational sample, we find a probability of 0.07 that they are drawn from the same distribution (i.e., that the observational SLF matches the theoretical SLF). An internal comparison between the Monte Carlo samples shows that only $5\%$ of the samples have probabilities equal to or less than 0.07. From these comparisons we find that the observational SLF is top-heavy compared to the theoretical expectations with discrepancy in the shape of the SLF at the 1.5-$\sigma$ level. Remember, however, that we are working with a small observational sample, as we surveyed only six hosts and covered only a small fraction of their virial volumes. Considering the good agreement between the theory and observations in the total number of dwarfs, we conclude that the observations match the theoretical expectations reasonably well. Further theoretical work after adding the LBT-SONG far-sample to our data will allow us to draw stronger conclusions. \par

\section{Conclusion} \label{section:conclusion}
In this work, we presented observational results from the LBT-SONG near-sample, which utilizes deep ($m\sim27$ mag) LBT/LBC images to search for dwarf satellites of six nearby, star-forming galaxies. We search for dwarfs as clustered point sources and characterize our detection efficiency using a combination of image simulations and a new catalog modelling technique which explicitly models the blending of stars. We show that the catalog modelling technique, after calibration to image simulations, serves as an excellent emulator for the detection efficiency with greatly reduced computational cost, and that treatment of unresolved blending is crucial for application of this method in the semi-resolved regime. We argue that similar analyses of detection efficiency should become standard in the field of luminosity function measurements, in particular exploration of the effects of both size and luminosity, as detection efficiency drops at fixed $M_{\text{V}}$ for increasing $r_h$. \par 

From our search, we find no new dwarf satellites, but we recover two previously known classical dwarf satellites. We present new photometric and morphological measurements of LV J1228+4358, an apparently disrupting dwarf of NGC 4449. Adding in DDO 113, a dwarf of NGC 4214 previously studied in \cite{Garling2020}, our observational sample of dwarf satellites consists of two classical dwarfs. Interestingly, both of these dwarfs show signs of environmental interactions with their hosts. The star formation in DDO 113 was quenched about 1 Gyr ago \citep{Weisz2011}, likely by strangulation \citep{Garling2020}, the cessation of cold gas inflows following accretion of the dwarf by the host, while LV J1228+4358 shows photometric irregularities most easily explained by tidal disruption \citep[e.g.,][]{Rich2012,Martinez-Delgado2012,Toloba2016a}. For both of our observed dwarfs to have such distinct signatures of host interactions indicates that sub-MW mass hosts may affect their satellite galaxies more strongly than previously thought. \par

Finally, we preview a new theoretical framework for predicting SLFs from theory and compute the mean expected SLF for the LBT-SONG near-sample. This phenomenological approach combines theoretical models for halo mass functions and subhalo mass functions with empirical models for the galaxy-halo connection and secondary galaxy properties to formulate a joint PDF that can be marginalized over to produce mean expected SLFs, and sampled to produce mock systems. We further correct for our photometric incompleteness using our dwarf detection efficiency measurements and our restricted field of view, producing expectation values that properly account for our observational limitations. We find that our theoretical expectation, in comparison to the observational sample, has one fewer classical dwarf and one more faint dwarf of $M_{\text{V}}\sim-7.5$ (i.e., the observational sample is top-heavy). However, overall agreement between the observational and theoretical results is reasonable given the small sample size. We expect to be able to draw stronger conclusions when we add the LBT-SONG far-sample into our theoretical analysis. \par

Our framework is well-suited to interpret observational results from current and future surveys, such as SAGA \citep{Geha2017,Mao2020} and MADCASH \citep{Carlin2016,Carlin2021}, which have and will continue to measure SLFs for large samples of hosts. The simulation-based work that is often done to compare the Milky Way's dwarf satellite population to theory \citep[e.g.,][]{Nadler2020} provides more detail than our method, but it is also much more computationally expensive and so calculating expectations for large ensembles of hosts is less viable. It is also difficult to vary the cosmology since it requires the simulation to be rerun. As our framework uses fully analytic models, it is highly extensible to alternate cosmologies and galaxy evolution models. A paper detailing this framework in full and exploring the impact of model choice is in preparation, which will follow the observational paper detailing the search for satellites in the LBT-SONG far-sample using integrated light detection methods. \par 

\hspace{\linewidth}
\section*{Acknowledgements}
C.G., C.S.K., and A.H.G.P. are supported by the NSF grant AST-1615838. C.G. and A.H.G.P. are additionally supported by the NSF grant AST-1813628. C.S.K. is also supported by NSF grants AST-1814440 and AST-1908570. D.J.S. acknowledges support from NSF grants AST-1821967 and AST-1813708. D.C. is supported by NSF grant AST-1814208. \par
The LBT is an international collaboration among institutions in the United States, Italy and Germany. LBT Corporation partners are: The University of Arizona on behalf of the Arizona Board of Regents; Istituto Nazionale di Astrofisica, Italy; LBT Beteiligungsgesellschaft, Germany, representing the Max-Planck Society, The Leibniz Institute for Astrophysics Potsdam, and Heidelberg University; The Ohio State University, representing OSU, University of Notre Dame, University of Minnesota and University of Virginia.\par
This research has made use of the NASA/IPAC Extragalactic Database (NED), which is operated by the Jet Propulsion Laboratory, California Institute of Technology, under contract with the National Aeronautics and Space Administration. This research has made use of NASA's Astrophysics Data System.

\section*{Data Availability}

The data underlying this article are proprietary but may be shared on reasonable request to the corresponding author.

\bibliographystyle{mnras}
\bibliography{library,annikarefs}
\bsp	
\label{lastpage}
\end{document}